# The Policy Implications of Economic Complexity

César A. Hidalgo

Center for Collective Learning, ANITI, IRIT, TSE, IAST, University of Toulouse, 21 All. de Brienne, 31000 Toulouse, France

Center for Collective Learning, CIAS, Corvinus University, Budapest, Közraktár u. 4-6, 1093, Hungary

Alliance Manchester Business School, University of Manchester, Booth St W, Manchester M15 6PB, United Kingdom

**Abstract**

In recent years economic complexity has grown into an active field of fundamental and applied research. Yet, despite important advances, the policy implications of economic complexity remain unclear or misunderstood. Here I organize the policy implications of economic complexity in a framework grounded on 4 *Ws*: *what* approaches, focused on identifying target activities and/or locations; *when* approaches, focused on timing support for related and unrelated activities; *where* approaches, focused on the geographic diffusion of knowledge; and *who* approaches, focused on the role played by agents of structural change. The goal of this paper is to provide a framework that groups, organizes, and clarifies the policy implications of economic complexity to facilitate its continued use in regional and international development.

1. Introduction

In less than two decades, economic complexity grew from a handful of papers into an active field of research (Hidalgo, 2021). Today, scholars and practitioners use economic complexity methods to explain variations in diversification patterns (Bustos et al., 2012; Hausmann et al., 2014; Hidalgo et al., 2007; Jara-Figueroa et al., 2018; Neffke et al., 2011;

Neffke and Henning, 2013), economic growth (Chávez et al., 2017; Doğan et al., 2022; Domini, 2019; Hausmann et al., 2014; Hidalgo and Hausmann, 2009; Koch, 2021; Lo Turco and Maggioni, 2020; Ourens, 2012; Stojkoski et al., 2016; Stojkoski and Kocarev, 2017), inequality of income and gender, (Barza et al., 2020; Basile and Cicerone, 2022; Ben Saâd and Assoumou-Ella, 2019; Chu and Hoang, 2020; Fawaz and Rahnama-Moghadamm, 2019; Hartmann et al., 2017; Sbardella et al., 2017), and sustainability (Can and Gozgor, 2017; Dong et al., 2020; Dordmond et al., 2020; Fraccascia et al., 2018; Hamwey et al., 2013; Lapatinas et al., 2019; Mealy and Teytelboym, 2020; Neagu, 2019; Romero and Gramkow, 2021; Sbardella et al., 2022) (Figure 1). This has made economic complexity methods increasingly common in policy reports and national development strategies (Balland et al., 2018; Hausmann et al., 2011; Mealy and Coyle, 2021; Montresor and Quatraro, 2019) that have been used to justify the creation of several data observatories by ministries of economy or production, or by national innovation or statistics agencies in Mexico, Chile, Brazil, Peru, and Estonia, among other places. But despite these advances, the policy implications of economic complexity are sometimes misunderstood. This is due in part to the rapid growth of the field, and also, to the fact that—as an interdisciplinary endeavor—economic complexity builds on network science and machine learning methods that are uncommon in economic geography, international development, and science, technology, and innovation studies. The goal of this paper is to help fill this gap by organizing attempts to bring economic complexity into practice in a framework that integrates multiple approaches.

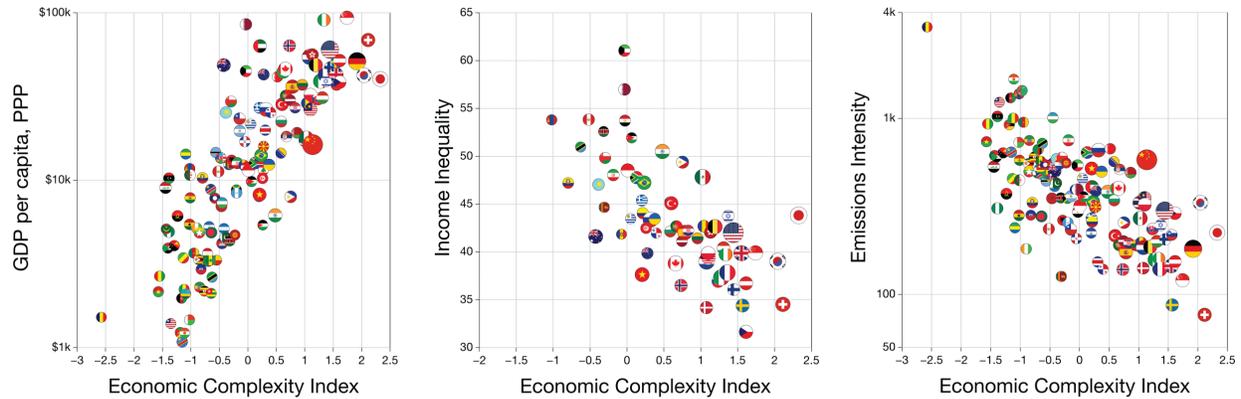

**Figure 1** *Economic complexity has been used to explain various macroeconomic outcomes, such as GDP per capita, income inequality, and emission intensities (emissions per unit of GDP per capita). GDP per capita data comes from the World Bank's WDI, inequality data from the Estimated Household Income Inequality (EHII) dataset, and emission intensity data shows greenhouse gas emissions (in kilotons of $CO_2$ equivalent, $CO_2e$) as a share of GDP (from XXX).*

Before diving into the framework, we need to define economic complexity, both as an academic field and as a collection of methods. In brief, economic complexity is the use of network science and machine learning techniques to explain, predict, and advice changes in economic structures. The focus on economic structure is motivated by work showing that these structures explain and predict important macroeconomic outcomes, from economic growth to the intensity of greenhouse gas emissions and income inequality (For a recent review see: (Hidalgo, 2021)). This work has helped formalize and expand intuitions that have for long been present in economic development, from Alexander Hamilton's Report on Manufactures, a document advocating for the industrial development of the United States (Hamilton, 1791), to more recent work on the importance of export structures for economic development (Hausmann et al., 2007; Imbs and Wacziarg, 2003; Rodrik, 2006; Saviotti and Frenken, 2008; Saviotti and Pyka, 2004).

But economic complexity is also a peculiar field, since it involves contributions from scholars from a wide range of disciplines, from the physicists and computer scientists who have pushed the development of machine learning and network science techniques, to the economic geographers, innovation scholars, and development economists that use economic complexity methods in their empirical work.

So, what explains the rise of this discipline?

Economic complexity techniques have become popular because of their ability to work with fine grained data in ways that preserve the identity of the elements involved and their patterns of interaction (Hidalgo, 2021). They provide a non-aggregative approach that allows a more nuanced understanding of economic structures without having to rely on coarse categories, such as changes from "agriculture" to "manufacturing" or from "manufacturing" to "services." This is an important departure from traditional economic theory, which for mathematical convenience, often assumes factor mobility across sectors (e.g. Ricardian trade theory and Heckscher-Ohlin). Economic complexity methods acknowledge the non-fungibility of factors such as knowledge, with limited mobility across sectors, which introduce path dependencies in the evolution of productive structures. While these ideas have been present in qualitative streams of literature, such as those in evolutionary economic geography (Boschma, 2005; Boschma and Frenken, 2011; Nelson et al., 2018; Nelson and Winter, 1985), economic complexity provides an empirical and quantitative counterpart to these qualitative theories that can be used to analyze the potential and impact of specific sectors in connection with formal models of economic growth from classical (Solow, 1956) and endogenous growth theory (Aghion and Howitt, 1992; Romer, 1990; Weitzman, 1998).

The ability of economic complexity methods to advance this non-aggregative view comes from two key concepts (Hidalgo, 2021): the idea of relatedness (Hidalgo et al., 2018, 2007), which is the use of recommender systems (Maes, 1995; Resnick and Varian, 1997)—a common machine learning method—to explain activities that a location will enter or exit in the future, and metrics of economic complexity (Hausmann et al., 2014; Hidalgo and Hausmann, 2009), which apply dimensionality reduction or spectral clustering techniques (e.g. PCA, SVD) (Mealy et al., 2019) to estimate the value of a country or region's specialization pattern (products (Hidalgo and Hausmann, 2009), industries (Chávez et al., 2017; Fritz and Manduca, 2021), patents (Balland et al., 2018), etc.).

Relatedness metrics help formalize the idea of path dependencies by anticipating the probability that a location will enter or exit an economic activity (Hidalgo et al., 2018, 2007). For instance, anticipating the probability that Austin, TX will increase its patenting activity in a specific patent class ("IPC G11C 5/14: power supply arrangements for information storage"). Relatedness has thus opened a more pragmatic approach to industrial policy where recommendations can be tailored to each activity and location. This has helped push development advice away from stereotypical global champions (e.g. A.I., green energy, biotech, etc.), which not all regions can compete in. Yet, this has also led to misinterpretations of the policy implications of economic complexity since, as we will see later, its implications are not to simply target related activities.

Complexity metrics estimate the value or sophistication of specialization patterns. This is validated by their ability to predict future economic growth (Chávez et al., 2017; Doğan et al., 2022; Domini, 2019; Hausmann et al., 2014; Hidalgo and Hausmann, 2009; Koch, 2021; Lo Turco and Maggioni, 2020; Ourens, 2012; Stojkoski et al., 2016), and explain geographic variations in inequality, (Barza et al., 2020; Basile and Cicerone, 2022; Ben Saâd and Assoumou-Ella, 2019; Chu and Hoang, 2020; Fawaz and Rahnama-Moghadamm, 2019; Hartmann et al., 2017), and emissions (Can and Gozgor, 2017; Dong et al., 2020; Dordmond et al., 2020; Fraccascia et al., 2018; Hamwey et al., 2013; Lapatinas et al., 2019; Mealy and Teytelboym, 2020; Neagu, 2019; Romero and Gramkow, 2021). This makes them useful in efforts of structural upgrading, especially when complemented with other metrics focused on environmental sustainability (Romero and Gramkow, 2021) or inequality reduction (Hartmann et al., 2017). These results have also been validated during the last decade in dozens of independent studies showing the applicability of these methods for a wide range of geographic scales (neighborhoods, municipalities, cities, regions, and countries) and activities, from urban amenities and research areas to patentable technologies and product exports (Balland and Rigby, 2017; Bandeira Morais et al., 2018; Barza et al., 2020; Boschma et al., 2015; Chávez et al., 2017; Chu and Hoang, 2020; De Waldemar and Poncet, 2013; Felipe et al., 2012b; Fritz and Manduca, 2021; Guevara et al., 2016; Hartmann et al., 2017; Hidalgo et al., 2018, 2020; Jara-Figueroa et al., 2018; Koch, 2021; Kogler et al., 2013; Lo Turco and Maggioni, 2020; Lyubimov et al., 2017, 2018; Ourens, 2012; Poncet and de Waldemar, 2013; Romero and Gramkow, 2021; Sbardella et al., 2017; Stojkoski et al., 2016; Zaldívar et al., 2019).

It is thus not surprising that economic complexity methods have enjoyed rapid adoption among policy practitioners. This could be due to several reasons.

First, they build and formalize intuitions that have for long been present in economic development theory.

The idea that the industrial fabric of an economy matters has been present since the dawn of industrial policy (Hamilton, 1791). During the twentieth century, this idea had multiple reincarnations, including the balanced growth theories of Paul Rosenstein-Rodan (Rosenstein-Rodan, 1961, 1943) and Walt Whitman Rostow (Rostow, 1959), the unbalanced growth theory of forward and backward linkages of Albert Hirschman (Bontadini and Savona, 2017; Hirschman, 1977), and the Prebisch-Singer hypothesis (Harvey et al., 2010; Prebisch, 1962; Singer, 2012), positing that developing countries specialized in commodities will experience deteriorating terms of trade. The idea is also related to the theories of development advanced by Alexander Gerschenkron (Gerschenkron, 2015, 1963) and Bela Balassa (Balassa, 1985, 1978).

Rosestein-Rodan advanced the idea that the social returns of an investment can be larger than its private returns when this investment occurs in the presence of complementary activities (Rosenstein-Rodan, 1943). This is related to the idea of economic multipliers and is used as an argument for "big push" approaches to economic development. Big push approaches propose investing simultaneously on multiple complementary industries, and thus, are known as "balanced" growth strategies. Rostow also emphasized the role of industry on development, but instead focused on a model based on stages of economic growth (Rostow, 1959). The model starts with traditional agricultural societies that take-off when they can shift resources from private households to professionalized economic activities. Eventually, economies industrialize, reducing their reliance on agriculture and moving into mass consumption models centered on a diversified industrial base. The last stage of Rostow's model is characterized by a search for quality, where households search for better quality instead of quantity of goods.

Hirschmann's work on forward and backward linkages is an example of an unbalanced growth strategy, since it emphasizes complementarities and spillovers along value chains, and thus, focuses on growing clusters of linked sectors (Hirschman, 1977). Finally, the Prebisch-Singer hypothesis (Prebisch, 1962; Singer, 2012) posits that developing countries specialized in the export of commodities will experience deteriorating terms of trade, and therefore, must industrialize to develop.

While these theories have important differences, they share a common emphasis on the importance of industrialization and on the need to consider complementarities and/or linkages between sectors (these ideas are also found, for instance, in the work of Michael

Porter on clusters (Porter, 1998, 1990) or more recent work by Hausmann and Rodrik (Hausmann et al., 2007; Rodrik, 2006), Chang, and Lin (Lin and Chang, 2009; Lin, 2011)). These ideas are also at the core of economic complexity research, albeit in a more mathematical and data driven form.

Yet, these similarities do not imply a symmetry in policy implications. For instance, Prebisch was an early and strong advocate of import substitution as a mean to shift imports from non-essential consumer goods to capital goods (Irwin, 2021). This is at odds with economic complexity research, which is more inline, for instance, with the work of Balassa (Balassa, 1985, 1978), the Hungarian-American economist who advocated for export promotion rather than import substitution. Economic complexity work is also different from Gerschenkron's theories of late comer advantage (Gerschenkron, 2015, 1963), which emphasized capital over knowledge by suggesting that late comers could jump into the productive frontier with larger plants and newer technology, if they could secure the finance (Freeman, 2002). But as it has been pointed out (Bell and Pavitt, 1993), this focus on capital and imported technology disregards the nuances of non-fungible knowledge accumulation that are key for specific sectors (e.g. Gerschenkron assumes that learning is relatively easy once the capital becomes available). As a conceptual framework, economic complexity deeply acknowledges the specificity of knowledge that limits its mobility across sectors (it is hard to transition cotton farmers to electronic chip makers, even in the long run). As such, it is a development theory more aligned with work focused on learning rather than capital accumulation (Argote, 2012; Arrow, 1971; Bell and Pavitt, 1993; Dasgupta and Stiglitz, 1988; Stiglitz, 2017).

In the context of this literature, economic complexity research represents an example of an unbalanced growth theory (e.g. à la Porter or Hirschman), since it provides methods to identify tailored diversification strategies based on a region's current pattern of specialization. This unbalance is justified by the limited mobility of factors across sectors, which is expressed in the intricacy of networks such as the product space (Hidalgo et al., 2007). But unlike Hirschman, which focused on value chains, economic complexity started with a more agnostic approach to economic linkages (Hidalgo et al., 2007), from shared knowledge to shared infrastructure and institutions (collectively described as "capabilities"(Hausmann and Hidalgo, 2011)). Soon, however, the literature on economic complexity begun to emphasize the importance of knowledge as a key input (Hausmann et al., 2014; Hidalgo, 2015). This is again, because its non-fungible nature limits its mobility across sectors and adds a combinatorial dimension explaining the path dependency of economic structures (Hidalgo, 2022). These specificities were explored empirically in work focused on unpacking-relatedness into multiple knowledge channels (Jara-Figueroa et al., 2018) or comparing the relative importance of knowledge agglomerations vis-à-vis value chain relationships (e.g. knowledge relations have increased in relative importance over time (Diodato et al., 2018)). This makes economic complexity a complement to endogenous growth theory (Aghion and Howitt, 1992; Romer, 1990), but instead of emphasizing knowledge's non-rival nature, economic complexity theory emphasizes its non-fungibility. These strong connections to streams of literature in economic growth and development have made economic complexity attractive to scholars and practitioners trained in these various traditions.

A second reason supporting the adoption of economic complexity methods in policy circles comes directly from the demand side of practitioners. Complexity approaches speak directly to a frustration that runs deep among many policymakers, especially in developing countries. These are policymakers who have grown tired of development advice that is either too unspecific (e.g. level the playing field, improve institutions, adopt best practices, etc.), or uniform, such as the economic liberalization package known as the Washington Consensus. The Washington Consensus started as a list of ten policy recommendations published by the economist John Williamson in an attempt to summarize ideas believed to be consensual among OECD countries, but different from standard practices in the developing world (Williamson, 2009). These ideas included *Fiscal Discipline*, *Liberalization of Interest Rates*, *Trade Liberalization*, and *Deregulation* (among others). The Washington Consensus also did not promote industrial upgrading, since Williamson did not consider it to be a determinant factor in the success of East Asian economies (Williamson, 2009). Thus, the recommendation was for economies to double-down on their current patterns of comparative advantage because it did not matter if a country specialized on iron ore or jet engines[*].

But the Washington consensus proved to be quite controversial. In fact, several practitioners, particularly in Latin America, Asia, the Middle East, and Africa, held on to the intuition that economic structures matter, and that doubling down on their economy's current patterns of comparative advantage (raw minerals and agriculture) was not the best long-term strategy. This frustration was accentuated by the fact that many of those who followed the advice, and engaged in the recommended institutional reforms, failed to reap the expected benefits, especially when they adopted reforms as a short-term signaling strategy (Andrews, 2013). This frustration was reinforced by a lack of sector specific approaches that could satisfy the demand of those looking to promote structural change. This ideological debate also partly explains some of the demand side for economic complexity methods, since they represent a middle-ground between some of these ideas. On the one hand, economic complexity methods are used to inform industrial policy, but on the other hand, they do so in the context of open economies engaged in export promotion instead of import substitution.

---

[*] A clear example of this argument can be found in Paul Krugman's 1993 AER paper" What do Undergrads Need to Know About Trade? (Krugman, 1993)

Yet, regardless of the reason for their adoption, today, economic complexity methods can be seen in reports focused on Smart Specialization in Europe (Balland et al., 2018; Deegan et al., 2021; Foray et al., 2009; Hassink and Gong, 2019; Montresor and Quatraro, 2019), China's special economic zones (De Waldemar and Poncet, 2013; Kahn et al., 2018; Zheng et al., 2016), Mexico's Smart Diversification strategy (Economía, 2021), policy briefs (Pugliese and Tacchella, 2020; Sbardella et al., 2022), or white papers calling to upgrade the manufacturing sector in the United States (Karsten, Jack, 2022). Today, it is not uncommon to hear experts across the world debate about the need to upgrade, sophisticate, or "complexify" an economy, which is how the ideas of economic complexity are communicated in the mainstream. We can find these concepts on reports and studies focused on the economies of China (Chen et al., 2017; Dong et al., 2022; Ferrarini and Scaramozzino, 2015; Gao et al., 2021; Gao and Zhou, 2018; Guo and He, 2017; He and Zhu, 2019; Zhu et al., 2020), Mexico (Chávez et al., 2017; Pérez Hernández et al., 2019; Zaldívar et al., 2019), Russia (Lyubimov et al., 2017, 2018), Brazil (Britto et al., 2016; Dordmond et al., 2020; Gala, 2017; Jara-Figueroa et al., 2018; Swart and Brinkmann, 2020), Uruguay (Ferreira-Coimbra and Vaillant, 2009), Turkey (Coskun et al., 2018; Erkan and Yildirimci, 2015; Hartmann, 2016), and Paraguay (González et al., 2018). These ideas are also in efforts focused on the economic structures of developed nations, such as the United States (Balland and Rigby, 2017; Boschma et al., 2015; Essletzbichler, 2015; Farinha et al., 2019; Fritz and Manduca, 2021; Lo Turco and Maggioni, 2020; Rigby, 2015), Canada (Wang and Turkina, 2020a, 2020b), Australia (Reynolds et al., 2018), Italy (Basile et al., 2019; Cicerone et al., 2020; Innocenti and Lazzeretti, 2019a, 2019b, 2017; Stafforte and Tamberi, 2012; Tullio and Giancarlo, 2020), and the United Kingdom (Bishop and Mateos-Garcia, 2019; Mealy and Coyle, 2021).

On a more modern recent context, economic complexity methods also provide a complement to mission-oriented policy approaches (Mazzucato, 2018; Savona, 2018), which aim to rally innovation across multiple sectors by focusing on concrete missions. Mission oriented policies have a Rostowian flavor, and while they acknowledge the need to consider absorptive capacity (Cohen and Levinthal, 1990), they lack a quantitative methodology to estimate the potential absorptive capacity of a location in a specific sector. Economic complexity methods provide such estimates, helping to ground mission-oriented policies in a world where electoral ambitions can push missions into unfeasible territories.

This adoption, however, does not mean that economic complexity methods are clearly understood. This can be explained in part by some key epistemological issues.

Traditional approaches to economic development focus on identifying specific causal factors, and then using them as potential levers in policy interventions (Kleinberg et al., 2015). Complexity approaches are based instead on variables that capture combinations of factors. A useful analogy here is to think of a propensity or risk score in medicine. Consider an indicator for a patient's propensity to heart disease. A risk score combines multiple factors (age, tobacco use, body weight, diet, sex, etc.) into a single numeric value. Some of these factors can potentially be intervened (e.g. diet, tobacco) while others cannot (e.g. age). The risk score can be used to predict the chances of heart disease, but, the score cannot be a causal factor, even though its components (age, diet, tobacco, etc.) can be. In a similar way, a relatedness or complexity index estimates a "risk" or "potential" to enter an activity, to experience economic growth, or to reduce inequality, coming from a combination of factors. But unlike in the case of heart disease, relatedness and complexity provide propensity scores for systems where the exact factors and their combinations are unknown.

Consider the principle of relatedness (Hidalgo et al., 2018). Relatedness contributes to the propensity that a location enters or exits an activity (Boschma et al., 2015; Guevara et al., 2016; Hidalgo et al., 2007; Jara-Figueroa et al., 2018; Neffke and Henning, 2013). Whether these activities collocate, or experience labor flows, because they share labor, knowledge, supply chains, or infrastructure, is not the point. Relatedness approximates their combined forces if this propensity expresses itself in a repeated pattern of colocation. That makes it flexible enough to capture demand side spillovers, like those experienced by neighborhood scale amenities (Hidalgo et al., 2020), or knowledge spillovers (Jara-Figueroa et al., 2018). The estimates work even when the balance between these forces change, since co-location or labor flow patterns will tend to adapt accordingly. Thus, the policy implications of economic complexity are not about increasing relatedness or complexity, as if they were single causal factors, but about using these metrics as strategic indicators for the propensity of a location to enter and exit specific economic activity. They belong to what some recent authors call policy prediction problems (Kleinberg et al., 2015). They are a path to mapping a development strategy rather than a specific lever for intervention.

Similarly, complexity metrics provide an estimate of the overall potential of an economic structure, based on information on the geographic distribution of economic activities. As such, they help anticipate economic growth or emissions, not by identifying a specific factor, but by estimating their combined presence.

Putting this epistemological discussion aside, it is safe to say that, despite the widespread adoption of economic complexity methods, there is a need to clarify how these methods should be used in practice. In fact, many efforts to put these methods into practice have been ad-hoc or build on naïve interpretations of the possible policy implications (e.g. focus on related activities).

In the following pages, I organize efforts to bring economic complexity ideas into practice in a framework that is explicit about the strength and shortcomings of different approaches. I organize this summary around four simple albeit fundamental questions: *what*, *when*, *where*, and *who*. The purpose of organizing these efforts into these four *Ws* is not to claim that these four questions are all encompassing, but to provide a compact and memorable structure that helps organize important parts of the literature while providing an easy to remember moniker for practitioners. We explicitly do not consider *how* questions (except in the discussion) in an effort to avoid some of the pitfalls from the early development literature, which often jumped into recommendations (e.g. big coordinated push, import substitution) in absence of a quantitative framework that could provide quantitative diagnostics and probabilistic forecasts.

*What* questions focus on targeting specific activities to satisfy a specific goal (growth, sustainability, inequality, etc.). *When* questions explore the idea of windows of opportunity, and strategies that combine targeting related and unrelated activities. *Where* questions focus on the geographical availability of knowledge. And *who* questions focus on the role played by various agents of structural change. Going forward, I call this the *4Ws* or the *W$^4$* framework for economic complexity policy.

## 2. The 4Ws
### 2.1 What

*What* efforts are the most common way to bring economic complexity ideas into practice. They focus on either identifying the activities that geographies could diversify into (Balland et al., 2018; Hausmann et al., 2014; Hidalgo et al., 2007), or the geographies that are most suitable to the development of an activity.

*What* approaches use relatedness metrics to recommend the activities that an economy should target and complexity metrics to assess the potential value of each entry or exit. These methods can also be expanded to include other target metrics, such as estimates of the income inequality associated to an activity (Hartmann et al., 2017) or their emission intensities (Romero and Gramkow, 2021) (Figures 2 d to f).

The standard way to implement *what* approaches is to use a relatedness-complexity diagram (Figure 2 a), introduced in the 2011 edition of the Atlas of Economic Complexity (Hausmann et al., 2014).† For an economy (e.g. a country, city, or region), a relatedness-complexity diagram plots the relatedness of that location to each activity in the $x$-axis and the complexity (or other metric of value) of each activity in the $y$-axis. Here we consider the case for locations, but a similar case can be made for activities (e.g. Figure 2 c), by plotting the locations that are most related to an activity and the complexity of each location.

---

† This diagram has also gained recent popularity thanks to work using it in the context of Europe's smart specialization strategy (Balland et al., 2018)

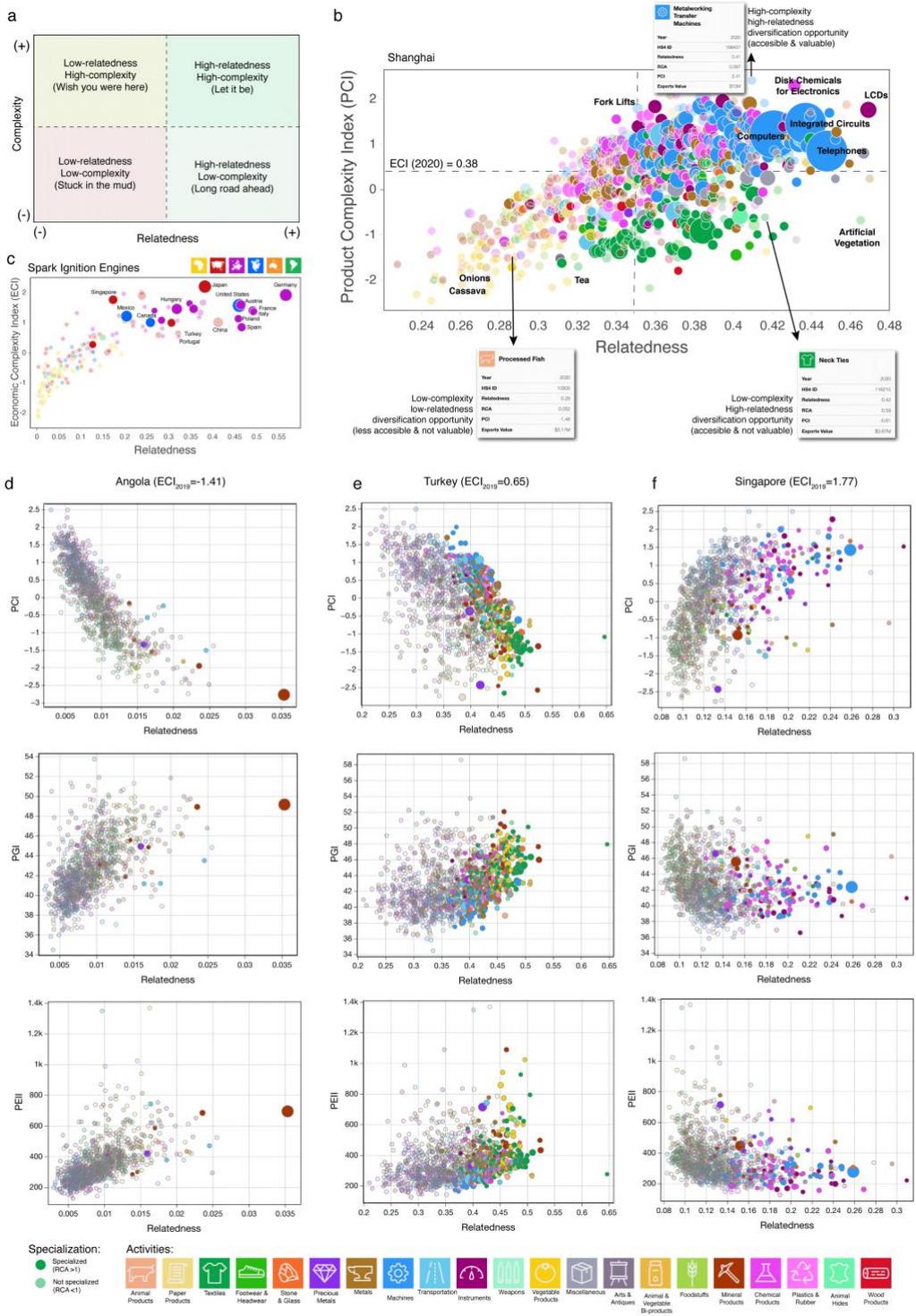

**Figure 2** What approaches. **a** schematic explanation of the quadrants in a diversification frontier or a relatedness-complexity diagram. Relatedness-complexity diagrams for **b** Shanghai exports, indicating potential export opportunities, and **c** spark ignition engines, indicating potential locations that could export them. Relatedness-complexity diagrams can be extended to other variables by replacing measures of product complexity (PCI) with measures of the inequality associated with a product (PGI (Hartmann et al.,

2017)) *or the expected level of emission intensity (PEII (Romero and Gramkow, 2021)). Here we present these diagrams for the economies of* **d** *Angola,* **e** *Turkey, and* **f** *Singapore.*

We can think about complexity-relatedness diagrams in terms of four quadrants (Figure 2 a). The top-right quadrant (high-complexity and high-relatedness) shows activities that are both desirable‡ (high-complexity) and accessible (high-relatedness). We call this the "*let it be*" quadrant, since it is a quadrant in which diversification is feasible and desirable. The top-left quadrant (high-complexity and low relatedness) shows activities that are desirable but less accessible. We call this the "*wish you were here*" quadrant, since diversification into activities in this quadrant is difficult yet desirable. The bottom-right quadrant (low-complexity and high relatedness) shows activities that are accessible but unattractive. We call this the "*long road ahead*" quadrant, since it shows activities where diversification is feasible, but that are not too attractive in terms of complexity. Finally, the bottom-left quadrant (low-complexity and low-relatedness) shows activities that are neither desirable nor accessible. We call this the *stuck in the mud* quadrant.

Figure 2 b uses export data from Shanghai to highlight products in different quadrants of the relatedness-complexity diagram. Products that Shanghai is exporting with comparative advantage (Shanghai is specialized in) are shown in full opacity. Products that Shanghai is not specialized in are shown with 50% transparency. Since Shanghai is a complex economy (ECI=0.38, shown by the dashed line), there are almost no products in the low relatedness high complexity quadrant. In fact, for Shanghai most products are in the high relatedness-high complexity quadrant. That means Shanghai is in an attractive structural position because its related diversification opportunities are also high complexity (and high income) activities.

---

‡ Beyond complexity, it is possible to use other desirability metrics associated, for instance, to a product's expected level of inequality (PGI) (Hartmann et al., 2017) or emissions (PEII) (Romero and Gramkow, 2021).

Using the hierarchical nature of product and industry classification systems to our advantage, we can generate these diagrams at various levels of aggregation. For instance, by presenting together garments products, such as neck ties and trousers, together as a single "garments" node. It is important to be careful with such aggregations, however, since product categories sometimes involve intuitive or unrelated groupings. For instance, at the HS4 level (a disaggregate level involving 1200+ products), products as different as espresso machines (HS6 841981) and medical laboratory equipment sterilizers (HS6 841920) are grouped into the same four-digit code "other heating machinery" (HS4 8419), highlighting the importance of disaggregate data.

When applied to activities (products, industries, technologies), the relatedness-complexity diagram can be used to recommend locations that are suitable for an activity. Figure 2 c shows a relatedness-complexity diagram for Spark Ignition Engines (using 2019 international trade data). The chart shows that spark ignition engines are a product related to Germany, France, and the United States, who are already specialized in them, but also, they are related to China and Turkey, who are not currently specialized in it (RCA<1). Thus, the diagram suggests that China and Turkey are two countries with a high potential to develop comparative advantage in spark ignition engines in the future.

Relatedness-complexity diagrams, sometimes referred to as the "diversification frontier," are an exploratory tool that people can use to identify diversification opportunities that are both feasible (high relatedness) and attractive (high complexity). This diagram can of course be modified by choosing different metrics of relatedness. For instance, we can use a model involving several predictors to get a measure of implied comparative advantage (Hausmann et al., 2021) or use machine learning approaches to estimate the probability of entry and exit (Tacchella et al., 2021). This is the approached used for instance in (Jun et al., 2019), which considers an extended gravity model together with three measures of bilateral relatedness to predict an economy's future exports to a destination. The approach can also be modified by using alternative metrics of complexity[§] or economic value, such as the market growth for an activity, its current market size, the income of current competitors, or measures of expected inequality (Hartmann et al., 2017) and emissions (Romero and Gramkow, 2021)[**]. In fact, starting with the seminal work of Hamwey (Hamwey et al., 2013), several papers have focused on "green diversification," by using these techniques to identify development paths that target green products or technologies (Dordmond et al., 2020; Fraccascia et al., 2018; Montresor and Quatraro, 2019; Moreno and Ocampo-Corrales, 2022; Ning and Guo, 2022; Perruchas et al., 2020;

---

[§] Including the Fitness index (Tacchella et al., 2012), the Activity index (Bustos and Yıldırım, 2022), GENEPY (Sciarra et al., 2020), MONEY (Gnecco et al., 2022), the innovation adjusted ECI (Lybbert and Xu, 2022), and value added corrected complexity (Koch, 2021). Some work comparing these different measures of complexity help us understand some of their similarity and differences. ECI and Fitness, for instance, have found to be highly correlated and in panel regressions provide a similar correlation with future economic growth. Using employment data for the United States (Fritz and Manduca, 2021) find a correlation between ECI and log Fitness of 0.94. Using international trade data (Hartmann et al., 2017) documents a correlation of 0.86 and (Bustos and Yıldırım, 2022) find a correlation of 0.87. In (Stojkoski et al., 2016), Fitness and ECI are compared in panel growth regressions, with the ECI regression explaining more variance in future economic growth than the Fitness regression (33.4% versus 29.7%). A similar exercise was conducted by (Bustos and Yıldırım, 2022) using baseline models in 5-year fixed-effects growth regressions. The Ability index performs the best in this case, explaining 13% of the variance, compared to 12% for ECI and 9% for Fitness (which is significant only at the 5% level). When it comes to inequality, a similar exercise was performed by (Hartmann et al., 2017), comparing the ability of ECI and Fitness to explain international variations in income inequality. At the cross-section that the ECI model explains 69.3% of the variance in inequality compared to the Fitness model (67.2%). They also show a larger effect in a fixed-effect model, where only ECI remains significant. Despite the high correlation between these metrics, there is work helping interpret some of these differences. Fitness metrics tend to be more correlated with traditional measures of diversity and/or concentration, whereas ECI metrics tend to capture aspects of composition that go beyond diversity. (Stojkoski et al., 2016) documents a correlation between Fitness and diversity of 0.89. Similarly (McNerney et al., 2023) finds a correlation between diversity and Fitness of 0.9, and groups Fitness metrics into a diversity type category and ECI metrics into a composition type category. This explains, for instance, why small, developed economies, such as Finland and Singapore, get high ECI scores, whereas larger economies, such as China and India, get higher Fitness scores.

[**] An example of a report using multiple targets can be found in a report prepared for the government of the Dominican Republic in 2011, which included several metrics of value, such as an activity's global market size, its growth potential, and whether the global market in an activity was growing or shrinking (Hausmann et al., 2011).

Santoalha and Boschma, 2020; Sbardella et al., 2022). Such modifications, however, do not change the basic idea behind the approach: identifying activities for potential upgrading that are feasible and attractive.

### 2.1.1. Using Relatedness-Complexity Diagrams in Practice

Using relatedness-complexity diagrams in practice requires understanding both the concepts behind these diagrams and the empirical patterns described by the data. While in principle, an activity or an economy can locate in any quadrant, in practice, the non-random nature of economic development implies some recurring patterns.

One well-known pattern is the reversal of the correlation between product complexity and relatedness observed for different levels of economic complexity (Figure 2 d)(Pinheiro et al., 2021). At relatively low levels of complexity (e.g. ECI < 0), economies exhibit a strong negative correlation between relatedness and complexity (as is the case for Angola in Figure 2 d). That means economies are more related to low complexity activities. In this case, the *let it be* quadrant is empty and the diagram is mostly populated in the other three quadrants. This pattern implies a tradeoff, since it means that for low complexity economies the most feasible activities are unattractive (low complexity) while the most attractive activities (high complexity) are hard to develop (low relatedness). This tradeoff can be explained by the fact that metrics of complexity are projections of matrices of similarity (Mealy et al., 2019). That means that low complexity products will tend to be near low complexity product. The result is to focus on a strategic frontier populated by the highest complexity activities for a given level of relatedness. But as we will see later, there are important considerations that escape what approaches and involve timing the strategic development of unrelated activities ("when" approaches). That is, looking not only on the first-order desirability of a product, but the second order desirability implied by increased access to neighboring activities.

For more complex economies, the negative correlation between complexity and relatedness weakens and eventually reverses (case of Turkey & Singapore in Figure 2 e and f). At the reversal point, economies are more related to complex activities (case of Singapore in Figure 2 f). Advanced economies, therefore, are in a much more favorable strategic position, since for them, the most attractive activities are also those that are most feasible. For these economies, the frontier is not one of identifying compromises, but one of rapid catchup or innovation. This analysis can be complemented by looking simultaneously at various metrics, such as the expected emissions and inequality associated with a product.

The observation that more complex economies are in a more favorable position may seem intuitive, or even naïve, until we notice that complexity and income are not perfectly correlated. In fact, it is the mismatch between complexity and income what help us predict future economic growth (Hidalgo and Hausmann, 2009). Consider Peru and South Korea. Back in 1973, Peru had an income per capita that was twice that of South Korea, and about four times the capital per worker. It also had a similar level of education (measured in years of schooling). But in 1973 South Korea was a more complex economy than Peru. Recall that ECI is measured relatively to the world average, so an ECI=0 means that a country is on the average of the world, and an ECI of 1 means a country is one standard deviation to the right of the world average. In 1973 South Korea had an ECI=0.86 and Peru had an ECI=-0.8, meaning that South Korea was 1.66 standard deviations more complex than Peru. Thus, back in 1973, complexity tells us that South Korea was in a better position for subsequent structural change and economic growth than Peru, even when it had half the income and a quarter of the capital per worker. This would be a hard conclusion to make using traditional aggregates (which are blind to the structural information available in the specialization matrices used to estimate relatedness and complexity).

The reversal of the correlation between relatedness and complexity has been proposed as an explanation for middle-income traps, since it is a pattern that differentiates between high- and low-complexity middle income economies. Those that succeed at crossing the chasm (e.g. Japan in the 70s, South Korea in the 80s and 90s, etc.) were relatively high complexity economies (at their level of income) compared to those that have not (e.g. Peru, Algeria, etc.).

Another empirical pattern that is also important to remember is the fact that relatedness is a stronger predictor of entries for low complexity economies (Pinheiro et al., 2021). This may wrongly lead us to conclude that low complexity economies should focus primarily on related activities, but as we will see next, this argument is flawed, since betting all development efforts on related activities can be shown to be a mathematically suboptimal diversification strategy. This is related to the idea that focusing too much on relatedness may overspecialize economies and risk technological lock-in (Boschma et al., 2012), trading off short-term adaptability for longer term evolvability. These fears of lock-in have pushed recent research to focus on unrelated diversification (Boschma and Capone, 2015; Zhu et al., 2019, 2017). But since unrelated diversification is empirically infrequent (Pinheiro et al., 2021), this research often finds statistical results that are weaker (smaller size effects), reducing rather than reversing the role of relatedness. Nevertheless, there are interesting findings. For instance, (Boschma and Capone, 2015) find that relatedness is a stronger predictor of diversification in economies with more coordinated forms of capitalism (but the size of the effect is about 1/10 to 1/3 of that of relatedness alone). Similarly, Zhu, Wang, and He (Zhu et al., 2019) find that the introduction of high-speed rail reduces the effects of relatedness, inducing more path-breaking development, but with a size effect that is again about 1/10 to 1/5 that of relatedness.

Overall, the moral of this section is that using relatedness-complexity diagrams in practice requires going beyond the theory. They involve knowing about the empirical regularities observed in these diagrams when applying them to multiple locations and/or activities.

### 2.1.2. Limitations of *What* approaches

*What* approaches attempt to identify diversification opportunities using measures of relatedness and complexity. While these approaches are expected to beat chance, they may still be somewhat naïve, since economic context is only partially captured by data on the geography of economic activities. Moreover, these approaches involve a classic example of the tension between positive and normative philosophy that often permeates policy discussions. What is "natural" (philosophically positive) for developing economies is to enter more related activities. But that may be undesirable (negative from a normative perspective), since following relatedness may lock-in these economies in low complexity activities. At the same time, what is desirable for these same economies—upgrading their productive structure—may be more difficult given the inertial force of relatedness. That's why the policy implications of economic complexity need to go beyond efforts to identify sectors to enter or locations to target.

*What* approaches epitomize the dream of a pragmatic, machine-like industrial policy. They are built on the idea of a "*neutral*" tool that avoids the *"biases"* of political influence. But it is also an approach that needs improvement. While the ability of relatedness to predict economies entering new product exports, industries, research areas, and patentable technologies, has been documented to the point of being acknowledged as a principle (Hidalgo et al., 2018), it is important to keep in mind that statistical significance can be large even when size effects are small. *What* approaches beat chance but are unable to remove it from the equation. They tell us about changes in probability, not certainties, and if used naively can lead us astray. The fact that economies enter related activities—on average—still means that they will fail many related diversification attempts. *What* approaches, therefore, are far from the end of the path. They provide at best a limited first glimpse of the promise of these methods. A promise that will largely remained unfulfilled, if we were to stop here. To move ahead, we need to flesh-out their limitations in search for complementary approaches.

3. When

*When* should an economy enter related or unrelated activities? How much should an economy invest in related and unrelated diversification? How should this calculation change as an economy climbs the complexity ladder? While *what* approaches focus on what activities to target, when approaches tell us about when to target related and unrelated activities.

*When* approaches were introduced recently by (Alshamsi et al., 2018) (expanded theoretically in (Waniek et al., 2020)). Unlike most work on relatedness, which tends to be statistical, Alshamsi's work starts by accepting relatedness as an empirical fact and building models to explore the question: what is the optimal strategy to diversify an economy? This brings into consideration the second-order desirability of entering a product, technology, or industry. The paths that it opens.

Using both stylized and numerical models, Alshamsi et al. (2018) show that, under relatively general conditions, strategies focused purely on relatedness (targeting the most related activity) are suboptimal diversification strategies. That is, they show that to diversify an economy faster you should not always target the most related activity. This result may seem counterintuitive, but it is also good news, since it provides both, a strong mathematical argument against the idea of focusing solely on relatedness, while opening a door to a portfolio-based view of complexity policy. In this portfolio view strategies look to balance efforts to enter related and unrelated activities.

We can get the intuition behind Alshamsi et al.'s 2018 main result in Figure 3 a. Consider a network of related activities (e.g. a product space, technology space, or industry space) that looks like a wheel, with a central node (a hub) and a ring of peripheral activities. Now consider an economy with a pattern of specialization on that network, represented by the darkened nodes. Since relatedness is an estimate of the chance that an economy enters a new activity, it is intuitive to think that diversification will be the fastest when we maximize the probability of success of each step (top path). That is, when we always pick the highest relatedness activity. The problem with this intuition is that relatedness is not a static quantity. In fact, changes in relatedness ripple through the network as an economy enters a new activity. So, thinking only one move ahead (picking the highest relatedness product) is not the optimal strategy. In fact, instead of targeting activities by simply maximizing relatedness, we should also consider the change in relatedness induced by each successful entry event (bottom path). This implies entering the central node in that network a bit earlier, when it is relatively unrelated (1/5), because unlike the nodes in the periphery, the central node accelerates subsequent diversification events. Thinking a few moves ahead changes the strategy to one where it is desirable to target some relatively unrelated activities if they enhance future diversification events. These are activities that link distant parts of the product space, research space, or technology space.

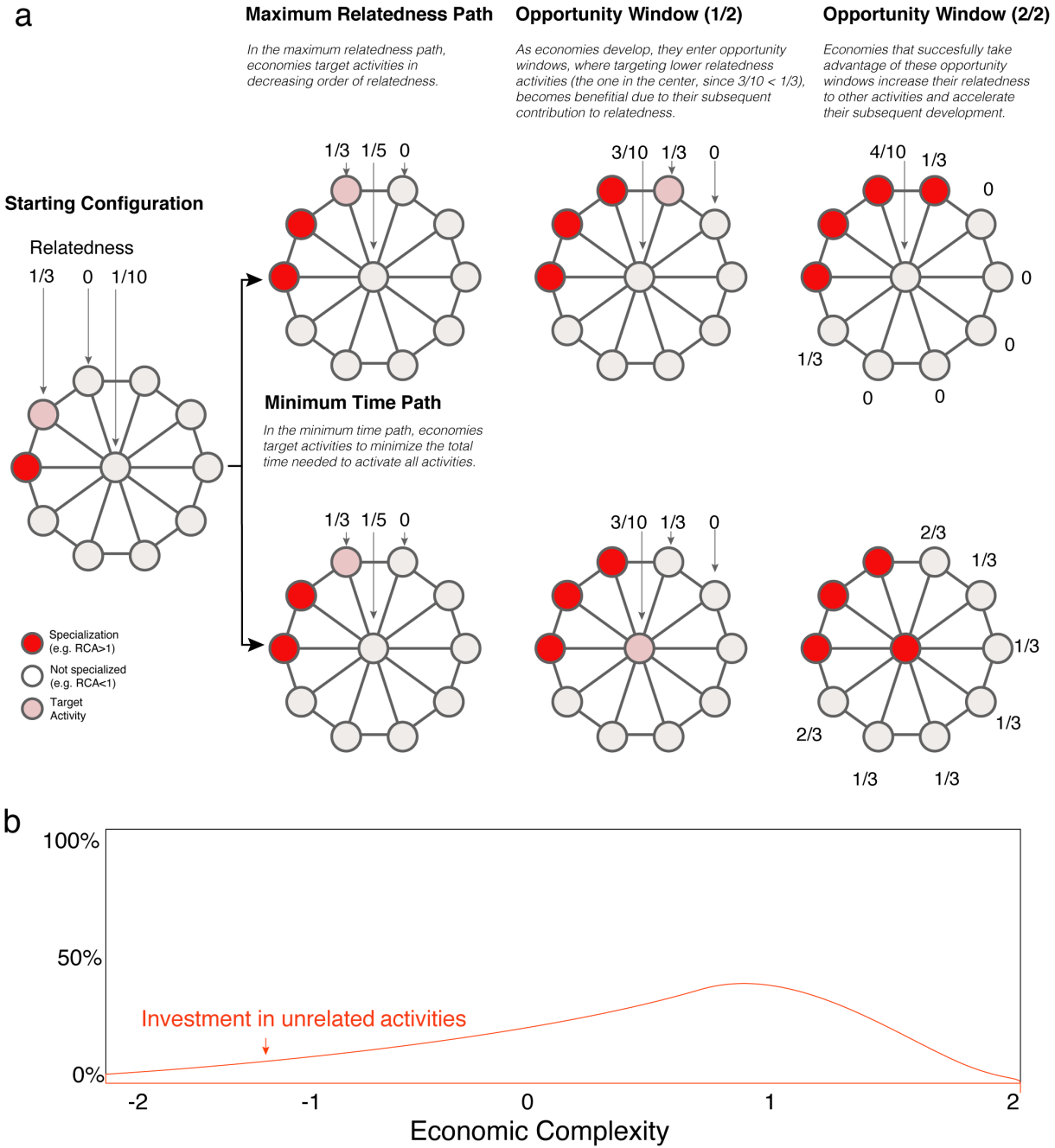

Figure 3. When approaches focus on balancing attempts to enter related and unrelated activities. a Illustration of the toy-model used by Alshamsi et al. 2018 showing that diversification strategies focused on maximizing relatedness can be mathematically suboptimal. b Illustrative development strategy that varies the level of investment in related and unrelated depending on an economy's level of economic complexity.

This result has three important implications. First, it pushes us to think about relatedness not only in terms of its present value, but in terms of its time derivative. Second, it also implies key timing considerations, providing the motivation for strategies focused on *when* to target related and unrelated activities. Finally, it also implies the need for a portfolio-based thinking, with different levels of support for related and unrelated activities.

The timing consideration relates to the ideas of opportunity windows and leapfrogging (Lee and Malerba, 2017; Malerba and Lee, 2021), since targeting an unrelated activity is only beneficial during a relatively narrow fraction of the development process. Targeting an unrelated activity too early, may lead to failed attempts and wasted resources. But targeting an unrelated activity too late, misses the opportunity to benefit from its spillovers (or from a short-lived cost advantage).

This relates to the need for a portfolio-based strategy focused on balancing support for related and unrelated activities.

For instance, consider a national innovation agency that splits its budget into "two" buckets[††]. One bucket is for supporting the development of related activities, which can be managed using *"what"* approaches. These are activities that the economy should be in a better position to enter. The second bucket is reserved for investing on more unrelated diversification attempts. Yet, these riskier investments are not random. They involve activities that can act as hubs connecting more distant parts of the product, industry, or technology space.

"When" approaches also tell us that the relative size of these two buckets is dynamic. In fact, their relative size should adapt to an economy's level of complexity (Figure 3 b). The size of the bucket for unrelated activities should be relatively large at an intermediate level of development, when the window of opportunity to enter unrelated activities is open and incomes are relatively low. Missing that window of opportunity could leave an economy with a relatively high income, but in a poor structural position. That is a challenging combination, since it represents a state of relatively low capacity without a price advantage.

---

[††] In fact, the explanation is easier assuming two buckets, but it is technically continuous.

Today, *what* strategies dominate the current use of complexity methods in policy, but *when* strategies may be the key for economies stuck in the middle-income trap (Bank, 2017; Felipe et al., 2012a). *When* approaches imply that the development model that may have brought these economies to middle income status may fail to push them past the chasm. In fact, it is countries with relatively high income and low complexity that are at risk of getting stuck at middle income and high inequality.

There is also some empirical evidence supporting the idea that targeting the most related activities may be suboptimal, and hence, supporting the general idea of "when" approaches. Using data on European projects, (Uhlbach et al., 2017) finds that, compared to unfunded projects, funding contributes more to the probability that a region will enter a technology for regions that are intermediately related to the technology. Regions that are already related to the technology enter with a similar probability whether funded or not. Regions that are too unrelated fail even with funding.

Still, despite these insights, compared to "what" approaches, "when" approaches seem relatively underdeveloped. In fact, to the best of our knowledge, determining the exact shape of the timing curve shown in figure 3 b is still an open question. While this shape can be determined analytically in trivial cases, such as highly symmetric toy models (Alshamsi et al., 2018), determining it for more complex topologies is a more challenging numerical exercise, and technically, and NP-complete problem (Waniek et al., 2020).

To make these jumps, countries, cities, and regions need to source knowledge. So, the next approach focuses on how knowledge moves across space, and how communication and transportation technologies can help "bend" space. We call these *where* approaches.

4. Where

One of the best-known facts about economic geography is the notion that knowledge diffusion decays with geographic distance. Unlike the relatedness and complexity approaches that emerged in the late 2000s (Hidalgo et al., 2007; Hidalgo and Hausmann, 2009), this fact can be traced back to the 80s and 90s (Jaffe, 1989, 1986; Jaffe et al., 1993), as it developed in parallel with the knowledge turn in economic growth theory (Aghion and Howitt, 1992; Romer, 1990, 1986). At first, researchers used patent citation data to show that knowledge spillovers decayed with geographic distance (Audretsch and Feldman, 2004, 1996; Jaffe et al., 1993). Subsequent literature provided a relational turn, by showing that this decay was explained primarily by the localization of social networks (Breschi and Lissoni, 2004; Singh, 2005). Eventually, scholars realized that other forms of proximity (Boschma, 2005; Torre and Rallet, 2005), beyond physical distance and social networks, could enhance or hinder knowledge diffusion. These calls for extensions were answered by the introduction of measures of relatedness (Hidalgo et al., 2007) capturing the "cognitive distance" between locations and activities‡‡, complementing the measures of geographic, cultural, and social proximity that dominated the literature until then.§§

But geographic, social, and cultural distance are still key factors shaping the diffusion of knowledge and the spatial concentration of complex (Balland et al., 2020) and innovative activities (Audretsch and Feldman, 1996). This means that policy efforts must consider physical geography and cultural factors when thinking about the policy implications of economic complexity.

This brings us to *where* efforts. These are efforts leveraging the opportunities implied by geographical proximity. The idea of learning from neighbors.

A historical example of the role of physical distance in the diffusion of technology is the invention of printing (Dittmar, 2011; Eisenstein, 1980; Innis, 2008; Jara-Figueroa et al., 2019; Pettegree, 2010). Removable type printing is an early example of a lucrative and complex technology that diffused from a source location (Mainz, Germany). The diffusion of printing was fast and characterized by intense competition (Pettegree, 2010). In fact, the market for printers saturated quickly, reaching a stable number of printers per capita in only 50 years (Jara-Figueroa et al., 2019). Places that were closer to Mainz, however, had an exogenous advantage that allowed them to adopt printing earlier. That meant that Paris was more likely to adopt printing earlier than Lisbon because it was "lucky" to be closer to Mainz. This provides a valid source of exogenous variation to create an instrumental variable that can be used to study the impact of printing (Dittmar, 2011; Jara-

---

‡‡ The cognitive distance interpretation of proximity and relatedness came later, in (Hausmann et al., 2014; Hidalgo, 2015, 2022)

§§ This is related to work on co-agglomerations (Ellison et al., 2010) and to recent efforts to unpack relatedness or co-agglomerations patterns showing the importance of related knowledge (Diodato et al., 2018; Jara-Figueroa et al., 2018).

Figueroa et al., 2019). In fact, research using this instrumental variable has shown that places that were closer to Mainz, and thus adopted printing earlier, begot famous scientists or artist—two cultural categories that rose with printing—sooner than places that were further from Mainz (Jara-Figueroa et al., 2019). This is a clear example of the role of geography on structural change.

Today, we can observe similar effects in work showing that countries and regions are more likely to enter the economic activities that are present in their geographic neighbors (Bahar et al., 2014; Boschma et al., 2013; Jun et al., 2019). This is an interesting fact when we consider that trade forces should push neighbors to differentiate. Even though countries trade more with their geographic neighbors (gravity effects), the effect of knowledge spillovers seems to be strong enough to sometimes overwhelm the differentiating forces of trade, especially on more technologically advanced sectors (Bahar et al., 2014; Jun et al., 2019). From a policy perspective, this observation invites us to consider geographic "gradients in economic complexity" as a potential for knowledge diffusion. That is, borders separating high and low complexity regions or countries can provide an opportunity for learning.

To grasp onto this intuition, consider Mexico and Cameroon. As a neighbor of the United States, Mexico has benefited from a strong gradient in complexity, and has grown the sophistication of its economy by becoming increasingly integrated with value chains in the United States. Northern states, such as Nuevo León, have benefited from this integration and today enjoy economies with a strong manufacturing base on sophisticated products (Figure 4). Yet, Nuevo León's exports go primarily to its neighbor to the north (85.2% go to the United States). This integration, has not reached the south of Mexico, where states like Guerrero suffer from low complexity and high poverty (27% of population in extreme poverty compared to 1.5% in Nuevo León ("DataMéxico," 2020)).

The result is that northern states have been pulling up Mexico's economic complexity and development, helping transform the Mexican economy into a manufacturing hub (Figure 4). As a matter of fact, in 2020 Mexico exported nearly as many cars as the United States ($41.6B vs $47.6B), and more cars than South Korea ($36.9.8B), Canada ($31.8B), or France ($18.9B). This is clearly a reflection of value chain integration, since Mexico is a net importer of key components, such as transmissions, combustion engines, and engines mounted on a chassis. In that same year, Mexico was also the second global exporter of video displays (after China), but a net importer of LCDs. This structural transformation did not happen overnight, but over decades, fueled in part, by the gradient of complexity between Mexico and the United States.

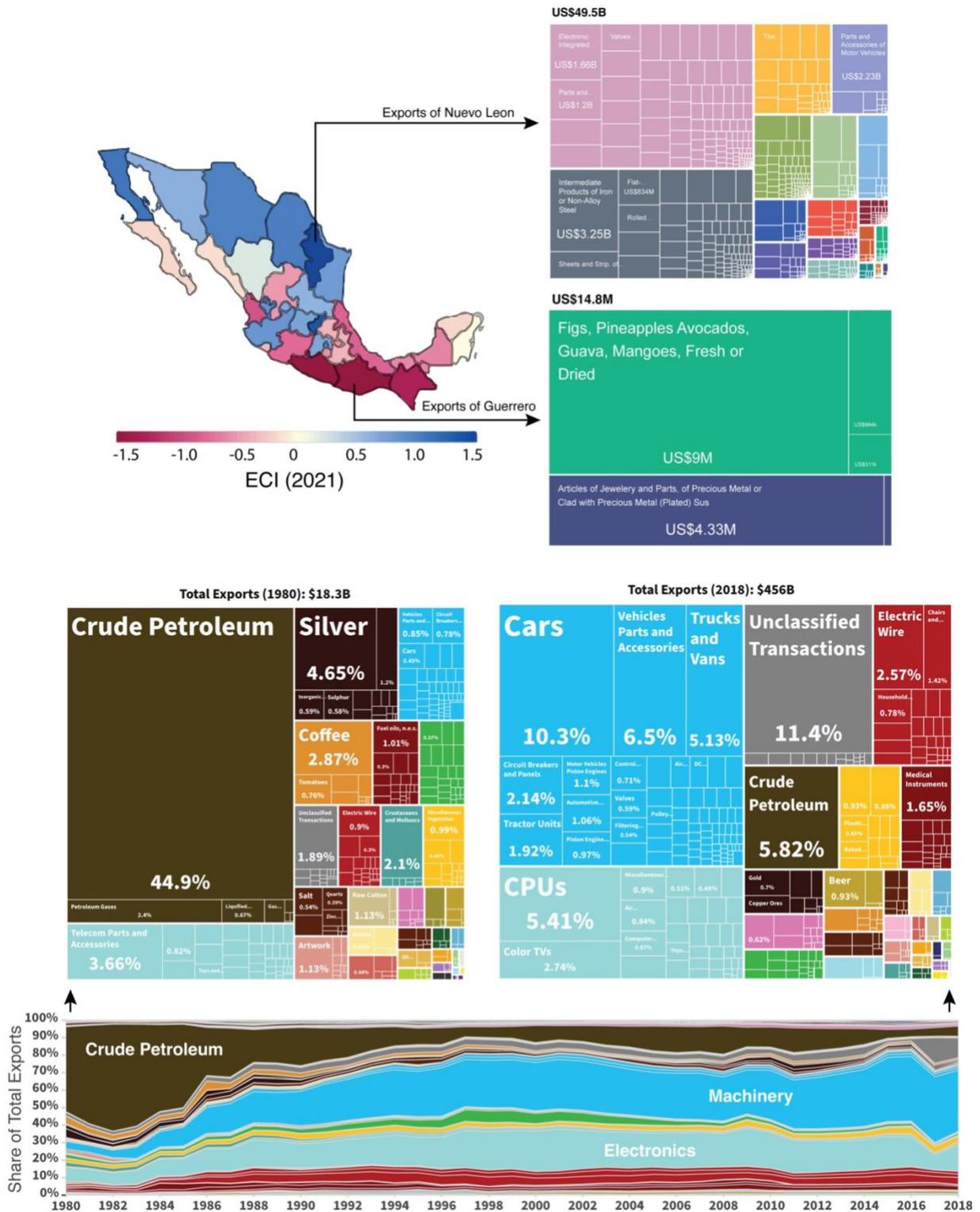

*Figure 4.* Economic complexity of Mexican states (from datamexico.org). Evolution of Mexico's export structure between 1980 and 2010 (from oec.world).

Now consider the case of Cameroon, a country which unlike Mexico, lacks high complexity geographic neighbors. This means that for a country like Cameroon, absorbing knowledge by integrating with the value chains of its neighbors is a more limited development strategy.[***]

Location matters. Countries like Mexico or Czechia are at an advantage when it comes to learning form their neighbors. By following *where* approaches, these countries can integrate with their neighbors and reap the benefits of their complexity. This is in fact a classic development idea that can be found in the *flying geese model* (Kojima, 2000) of development touted for decades by east Asian economies. The basic idea is to copy the economic successes of your geographic neighbors by leveraging their knowledge, physical proximity, and historical and cultural similarities. But where approaches can also go beyond the geography of production, and into the geography of trade destinations. In fact, extensions of the idea of relatedness to bilateral data (Jun et al., 2019) have shown that the effects of geography extend not only to what you export, but to where you export too.

So how can countries and regions increase their integration with their neighbors? Here is where we can find a few policy levers, such as changes in transportation and improvements in language skills and communication technologies.

Recent literature in economic geography has shown that improvements in transportation can accelerate geographic spillovers. For instance, work leveraging the expansion of high-speed rail in China showed a larger impact of rail on the industries shared by the provinces connected by high-speed rail (Gao et al., 2021). This suggests that faster rail connections promote knowledge diffusion and learning. Similarly, historical data from Sweden has been used to show that the introduction of trains in the eighteenth century accelerated the growth of cities located on the shortest paths between large urban centers (and for which the introduction of rail can be considered exogenous)(Berger and Enflo, 2017). This agrees with research using the introduction of discount flights and changes in travel rates to document increases in scientific co-authorships and patenting rates among the connected cities (Catalini et al., 2020; Hovhannisyan and Keller, 2015). Overall, these studies suggest that reductions in travel time promote geographic spillovers (by effectively reducing physical distance).

---

[***] When considering diversification through value-chains, is important to keep in mind that evidence points to backward/upstream linkages as the key direction of diversification (Bahar et al., 2019; Bontadini and Savona, 2017; López González et al., 2019).

There is also evidence showing that the introduction of communication technologies can "reshape geography." Research using data on the gradual arrival of internet to Africa has documented a positive effect on employment, especially in high skilled occupations (Hjort and Poulsen, 2019). Similarly, using data on internet penetration in China, researchers have shown that internet rollouts boosted firm manufacturing exports (Fernandes et al., 2019). When it comes to economic complexity, a study using the number of secure servers and a civil liberty index to instrument for internet adoption showed a positive effect of internet usage in economic complexity (Lapatinas, 2019). But communication technologies can be about more than digital connectivity. For example, research studying the effects of machine translation in an international trade platform estimated that machine translation resulted in a 10% increase in the exports mediated through that platform (Brynjolfsson et al., 2019).

Whether physical or digital, geography matters. *Where* approaches are about understanding the geographic constraints, and opportunities, implied by the spatial position of an economy. This involves leveraging within and between country learning opportunities, through cooperation, transportation, and communication technologies. While these technologies can "reshape" space, their strength may still not be quite enough to "fold it," inviting us to think strategically about how to use them in combination with other approaches.

## 5. Who

*Who* approaches focus on the agents of structural change (Elekes et al., 2019; Neffke et al., 2018). These are the people, organizations, or institutions that lead or catalyze efforts of structural upgrading, by either bringing knowledge to a region or setting up the ecosystem required for such upgrading to take place (Uyarra and Flanagan, 2021). This is of course, related to the long literature in migration, innovation, and trade (Brücker et al., 2012; Caviggioli et al., 2020; Kerr, 2018, 2008; Mayr and Peri, 2008; Miguelez and Morrison, 2022; Miguelez and Noumedem Temgoua, 2020; Parsons and Vézina, 2017; Rapoport, 2016; Saxenian, 2007) and to literature on innovation and institutions (Freeman, 2002; Nelson and Nelson, 2002).

An important branch of the former literature focuses on the role of foreign or non-local actors (e.g. migrants and foreign firms) on a location's ability to enter unrelated activities. In fact, there is a vast literature documenting these claims.

Recently (Elekes et al., 2019) used Hungarian manufacturing data to show that foreign owned firms deviate more than local firms from a region's current pattern of specialization. Similarly, (Neffke et al., 2018) used matched employer-employee data to show that "incumbents mainly reinforce a region's current specialization [and that] unrelated diversification […] originates [mostly] via new establishments […] with nonlocal roots." Lo Turco and Maggioni (Lo Turco and Maggioni, 2019) used data on Turkish manufacturing firms to show that entries into new actives are more related to the product mix of foreign firms than to that of domestic firms or local imports. Crescenzi et al (Crescenzi et al., 2020) used patent data to show that foreign multinationals have a positive effect on innovation rates, but that this effect is smaller for technology leaders, since these tend to engage in fewer alliances with local firms. Miguelez and Morrison (Miguelez and Morrison, 2022) studied migrant inventors in European regions to show that they contribute to unrelated diversification through knowledge creation and transfer. These findings are related to work showing that migrant inventors bring new knowledge to the regions they migrate into (Bahar et al., 2020; Breschi et al., 2017; Caviggioli et al., 2020; Kerr, 2008; Miguelez and Moreno, 2018; Miguelez and Noumedem Temgoua, 2020; Parsons and Vézina, 2017), to work showing that business travel promotes industries from the home countries of travelers (Coscia et al., 2020), and to literature emphasizing the role of foreign direct investment and technology transfer in structural upgrading and development (Lee and Lim, 2001; Mu and Lee, 2005). Overall, this research shows that foreign actors represent an important force that is complementary to *what* approaches and that promotes less related diversification.

The role of migrants as agents of structural change is explained in part by the skill bias of migrants. In the Gift of Global Talent, Bill Kerr documents this bias extensively, through various statistic (Kerr, 2018). Some are population level statistics, such as the fact that 5.4 percent of college educated workers migrate compared to 1.8 percent of those with high-school diplomas. Others, focus on the right-tail of the skill distribution, such as the fact that 10 percent of inventors with patents migrate, and that 31 percent of Nobel prize winners are migrants (a fraction that grew to 65 percent since 1970). This raises the question: is migration good news for everyone? Or only for the host countries and regions that receive talented migrants?

The intuitive reaction to migration is that it is beneficial to the host, but less so, to the country of origin. An effect known as "brain drain". Yet, this argument is debatable. On the one hand, migration has long stopped being a one-way trip. In fact, starting late in the 19[th] century, return migration became significant (Gmelch, 1980). On the other hand, there are several channels by which migrants can contribute back to their home territories. Beyond return migration, there is entrepreneurship, brain circulation, and even the incentives generated by a failed intention to migrate.

Return migration is an important channel of "reverse brain drain." Migrants can bring back skills, knowledge, and connections that they would have been unable to accumulate at home (Breschi et al., 2018; Choudhury, 2016; Gmelch, 1980; Mayr and Peri, 2008). In a study of R&D firms in India, for instance, Choudhry (Choudhury, 2016) uses exogenous variation in the assignment of managers to R&D units to show that return migrants have a positive effect in the patenting activity of the units they manage.

Another channel that benefits home and host countries are international entrepreneurship. Anna-Lee Saxenian develops this concept in her book "The New Argonauts" (Saxenian, 2007). Inspired by migrant entrepreneurs in Silicon Valley, Saxenian argues that these frequent fliers, operating firms with offices in both, their home countries and the Bay Area, defy the brain drain model and suggest instead a brain circulation model. These multinational firms act as channels for knowledge diffusion into geographies that are otherwise more peripheral when it comes to innovation. This is of course related to the literature on innovation highlighting the importance of ethnic networks. Using patent data, Kerr (Kerr, 2008) finds that foreign researchers cite more U.S.

researchers from their same ethnic group (30 to 50 percent more) than researchers from other ethnicities (in the same technology class). Breschi et al. (Breschi et al., 2017) find a similar effect for Asian inventors, but not for European inventors (except for Russians).

Finally, there is the idea that "migration prospects may foster human capital formation" (Beine et al., 2008; Rapoport, 2004). This is the idea that, in the presence of opportunities to migrate, people who would have otherwise not continued their studies, decides to deepen their education, even if later they do not end up migrating.

A separate strand of literature has explored the role of institutions in structural change and innovation (Chang, 1994; Freeman, 2002; Nelson et al., 2018; Nelson and Nelson, 2002). For instance, (Nelson and Nelson, 2002) argue that the idea of routines (Nelson and Winter, 1985) is tightly connected to institutions and innovation. Routines are programmatic sets of actions performed by the individuals in a firm, such as the routines used to decide on hiring and promotion or the steps followed by a sales team when interacting with a client. When passed on, these routines represent a form of culture or micro-institution. This also applies also to larger routines such as those that emerged, for instance, as innovations during the second industrial revolution. As firms became larger, and family groups could no longer provide the managerial talent and the capital needed, firms began to look for external sources of capital and professionalized management (Nelson and Nelson, 2002). Thus, changes in technology (e.g. railroads, electricity) led to changes in firm size distribution that in turn led to innovations in banking and managerial structure. Freeman (2002) takes a more historical approach when describing national systems of innovation, characterizing them through qualitative and quantitative measures, such as "strong links between scientists and entrepreneurs" and "gradual extension of primary, secondary, and tertiary education." But lacking a statistical counterpart, or narrowly defined quantitative models and definitions, it is hard to conclude whether these descriptions provide a true testable model of the role of institutions in structural transformation and change or a set of qualitative historical observations.

This is why more recent quantitative approaches, such as (Boschma and Capone, 2015) or (Tóth and Lengyel, 2021), are welcome. In Boschma and Capone 2015, the authors explore the interaction between relatedness and coordinated and uncoordinated forms of capitalism, finding a weak but significant effect of uncoordinated form of capitalism on unrelated diversification. A quantitative strand that would be related to some of the ideas proposed in the qualitative literature (Freeman, 2002; Nelson and Nelson, 2002; Nelson and Winter, 1985) is the work of (Tóth and Lengyel, 2021). By focusing on the structure of the social networks of innovators they get to something that is somehow related to routines and to some of the qualitative descriptors used by Freeman. In fact, (Tóth and Lengyel, 2021) find that firms are more likely to develop high-impact innovations when they hire inventors with a diverse network. Similarly, (van der Wouden and Rigby, 2019) find that "inventors in specialized cities value spatial proximity less and cognitive proximity more than inventors in diversified cities [since the latter can] partner with non-local inventors."

Finally, there are also case studies describing efforts by local actors to catalyze an environment for innovation. For instance, (Uyarra and Flanagan, 2021) interviewed 21 regional actors in Galicia, Spain, in work looking to understand a local effort to develop the unmanned aerial vehicle (UAV) industry. The case documents a coordination effort where local actors try to combine local assets, such as a disused military airfield, while attempting to attract market leaders into the region.

Overall, this case suggests that local actors behave in ways that are compatible with the findings of the quantitative literature, by working to attract non-local knowledge in efforts to promote diversification into unrelated activities.

## 6. Discussion: The 4Ws

Today, economic complexity tools are popular in economic geography, international development, and innovation studies. Yet, their policy implications are sometimes misunderstood. One reason behind this is that many research efforts have focused on either the, *what*, *when*, *where*, and *who* of structural change (mostly on the *what*). Without putting these four ideas together, the implications of economic complexity ideas can remain unclear.

This lack of cohesion can lead to misunderstandings. On the one hand, there has been work equating the policy implications of complexity to solely *what* efforts (the identification of related activities efforts) (Balland et al., 2018; Deegan et al., 2021; Hausmann et al., 2014), often ignoring the importance of (*when*) (Alshamsi et al., 2018), geographic proximity (*where*), or the role played by different agents in the process of structural change (*who*). On the other hand, there is qualitative literature that has been more critical of the quantitative work (Hassink and Gong, 2019; MacKinnon et al., 2019; Uyarra and Flanagan, 2021), but has nevertheless struggled to propose concrete alternatives while building on a relatively narrow understanding of the quantitative literature. In some ways, these misunderstandings may be an embarrassment of riches, since there are many research papers focused on single-*Ws* that it is hard to see the forest for the tress.

But in other ways, these misunderstandings may represent another embodiment of the tension between positive and normative approaches. After all, relatedness and complexity tools provide positive descriptions of reality. They help quantify the "inertial forces" limiting structural change. But that does not mean that the recommendation is to follow the inertia. So, when these approaches are misconstrued as normative, they lead to recommendations that are not what the methods intend to provide.

A more constructive way to resolve this tension is to build on the duality that exists between scientific and professional fields, such as physics and engineering, or biology and medicine. Scientific fields tend to be positive. They are concerned about the way the world is (e.g. understand the principles of physics and biology). Professional fields tend to be more normative, aiming to shape the world according to human needs and desires (build flying machines and curing diseases).

Relatedness and complexity represent positive descriptions of reality. An economy may be related to an activity but that does not necessarily imply that the economy should attempt entering it. This is just like when physics tells us how gravity works, not to imply that flight is impossible, but to help us consider gravity's pull. Thus, the policy implications of economic complexity are not to double down on what may be seen as inertial tendencies for economies (entering related activities), but to try to shape economies while taking these constraints and path dependencies into consideration. This implies a duality between an applied field, focused on policy, and a basic field, focused on the principles of economic complexity.

In this framework, *what* approaches describe the lay of the land. Relatedness and complexity tell us if an economy has quick wins available, whether structural change is difficult for that economy, or whether there are some paths that may be easier to climb. Complexity gives us an idea about an economy's potential output and help us understand which paths can be considered an upgrade for an economy. But to move the needle one needs to invoke the other *W*s. One needs to "defy gravity," albeit with pragmatism and respect.

*When* approaches tell us that there are windows of opportunities that we need to be vigilant about, and about the importance of seizing the opportunity when it arrives. *Where* approaches invite us to look for opportunities among our geographic and cultural neighbors, but also, to be strategic about the development of infrastructure and about the location of knowledge intense activities. *Who* approaches help us bend the principle of relatedness even further, by telling us about the key role played by the unrelated knowledge of migrants and their ability to bring economies to places that go beyond where locals can take them. The first *W* is about what is inertial. The last three are about how to think about changing an economy's state of motion.

But *how* do we bring these ideas into practice? What are the policy levers of economic complexity? Can we complement the 4 *W*s with a *how*?

As a toolbox, economic complexity methods do not provide new levers, but can help us improve how we evaluate and target the levers that countries and regions may already have in operation. Most countries, for instance, have national initiatives to fund science, technology, and innovation. By combining *what*, *when*, *where*, and *who* approaches, these initiatives can be organized into portfolios of related and unrelated activities, with a portfolio balancing strategy informed by a country's current level of economic complexity. The tools of economic complexity can also help organize these portfolios, by classifying projects as related or unrelated for each location and calibrating expectations for the success of related and unrelated diversification attempts. National innovation plans can also get a better lay of the land, of where the country or region is, and what are the sectors they expect to enter in the next 5, 10, or 20 years.

The tools of economic complexity also tell us about the importance of non-local knowledge, but also, about the importance of bringing non-local knowledge to an active ecosystem. On the one hand, they support the importance of smooth immigration and work permits for high-skill migrants. Attracting talent is key in today's world. But it is not easy. Small blocks on the road, such as unnecessary bureaucracy, can tilt the balance against a location trying to attract a talented migrant with multiple options. On the other hand, these lessons emphasize the need to bring non-local actors into the best possible local ecosystems, avoiding the temptation to create white elephant projects, such as the Yachay city of technology in Ecuador, a "knowledge hub" built hours away from any of the country's few urban centers (Vega-Villa, 2017) or the Neom city in Saudi (Nereim, Vivian, 2022).

Economic complexity tools also invite us to think about transportation links in terms of learning (Gao et al., 2021). Train lines, aircraft routes, internet connections, and automated translation software are all able to "bend space," even if just a little. The collective learning benefits of these technologies, however, are rarely used to motivate public investment, but should be a key arguments in their long-term strategic development.

Of course, we must consider key interactions among the 4Ws, since their use may depend on the measure of desirability used to identify activities. Efforts focused on growth, inequality, and emissions, may draw from a similar methodology, but may end up reaching different recommendations, for instance, in cases where growth and emissions are in conflict.

Moreover, putting complexity ideas into practice may still be difficult, in part, because economic complexity is a data hungry methodology that requires reliable fine-grained information. Without such data, even the most basic diagnoses may go beyond the analytical capacities of many economic development offices. These analytical capacities, have motivated the creation of comprehensive economic data observatories, such as the one used today by Mexico's secretary of the economy, as part of their smart diversification strategy (https://datamexico.org), by Peru's *Instituto Tecnológico de la Producción* (https://data-peru.itp.gob.pe), or by Brazil's official support service for small businesses, Sebrae (https://datampe.sebrae.com.br)[†††]. These tools not only help provide a unified showcase for dozens of government datasets, but in the case of Mexico, have become a tool used by the secretary of the economy to train commercial attachés in embassies and local governments to support their foreign direct investment efforts. To succeed, however, these tools need to be combined with the fine-grained knowledge available to local actors. They are after all, one more tool in the development practitioner's toolbox.

Still, it is unclear whether these efforts will manage to reproduce the benefits of organic or "evolved" forms of diversification. Policy stimulus may lead to diversification that lasts only as long as the incentives are in place, masking the absence of long-lasting capabilities. Further research will be needed to understand the effects of organic vis-à-vis policy stimulated diversification.

There are also, other avenues of research that are worth mentioning. Since the introduction of the economic complexity index (Hidalgo and Hausmann, 2009), several teams have introduced variants to estimate complexity, such as the Fitness index (Tacchella et al., 2012), the Ability Index (Bustos and Yıldırım, 2022), the GENEPY algorithm (Sciarra et al., 2020), the MONEY algorithm (Gnecco et al., 2022), the innovation adjusted ECI (Lybbert and Xu, 2022), and the Value Added Complexity (Koch, 2021). All these approaches, however, are constrained by the information that is available in a single dataset (mostly, international trade data), which may fail to reflect the geography of other activities, such as research and innovation. This has motivated a multidimensional turn involving work combining data on patents, papers, and products to create measures of cross-relatedness (Catalán et al., 2020; Pugliese et al., 2019a), and more recently, multidimensional metrics of economic complexity (Stojkoski et al., 2022). The latter shows that measures of complexity based on patents can complement trade-based measures of complexity by helping explain additional variance in economic growth, inequality, and emissions. This multidimensional turn is important, because complexity metrics applied to exports data are unable to capture key variation in the knowledge stock of countries that is expressed in patents or research data.

---

[†††] The author is a co-founder of Datawheel, a company involved in the development of such platforms. Datawheel did not finance this research project and his affiliation with Datawheel did not influence nor motivate the creation of this paper. The purpose of mentioning these platforms is because they provide concrete and relevant real world examples of the use of economic complexity methods in government settings.

Another trends, is the exploration of finer spatial scales, such as neighborhoods (Hidalgo et al., 2020), and firms (Bruno et al., 2018; Pugliese et al., 2019b), the latter echoing well-known work on firm coherence (Teece et al., 1994), or skills (Alabdulkareem et al., 2018). Yet, it is still not clear how these measures of complexity should be combined across scales, from neighborhoods to nations, or from skills to industries. Further research is needed to explore multiscale aspects of economic complexity.

Yet, some of the most important questions that remain in the literature, such as questions about the role of institutions and the identification of policy levers, may escape these improvements in methodology. In fact, despite great efforts, understanding the role of institutions in structural transformation remains an important challenge. While there is agreement on the intuition that industrial policies played a role in the industrialization of some key east Asian economies, the inability to reproduce this success in other parts of the world means that the precise mechanisms remain misunderstood. Moreover, while in the 1980s and 1990s, there was an apparent consensus about the institutions that best accompanied economic development (Williamson, 2009), that consensus disappeared together with the success of China, a large economy that achieved long term growth and rapid technological development using a radically different institutional model. This may suggest a decoupling between the role of political and economic institutions. An end of the "end of history" (Fukuyama, 1989) which invites further work to understand the interactions between political institutions, economic institutions, and culture.

Still, there is much to learn when it comes to the policy implications of economic complexity. My hope is that by organizing some recent contributions into this framework we push forth a literature that can dig deeper into these approaches, their interactions, and extensions.

### Acknowledgements

We acknowledge the support of the Artificial and Natural Intelligence Institute of the University of Toulouse ANR-19-PI3A-0004 and the 101086712—LearnData— HORIZON-WIDERA-2022-TALENTS-01 financed by EUROPEAN RESEARCH EXECUTIVE AGENCY (REA). The author is a co-founder and current CIO (Chief Innovation Officer) of Datawheel, a company specialized in the creation of data distribution and visualization systems, including oec.world and datamexico.org.

# References

Aghion, P., Howitt, P., 1992. A Model of Growth Through Creative Destruction. Econometrica: Journal of the Econometric Society 323–351.

Alabdulkareem, A., Frank, M.R., Sun, L., AlShebli, B., Hidalgo, C., Rahwan, I., 2018. Unpacking the polarization of workplace skills. Science Advances 4, eaao6030. https://doi.org/10.1126/sciadv.aao6030

Alshamsi, A., Pinheiro, F.L., Hidalgo, C.A., 2018. Optimal diversification strategies in the networks of related products and of related research areas. Nature Communications 9, 1328. https://doi.org/10.1038/s41467-018-03740-9

Andrews, M., 2013. The limits of institutional reform in development: Changing rules for realistic solutions. Cambridge University Press.

Argote, L., 2012. Organizational Learning: Creating, Retaining and Transferring Knowledge. Springer Science & Business Media.

Arrow, K.J., 1971. The Economic Implications of Learning by Doing, in: Hahn, F.H. (Ed.), Readings in the Theory of Growth: A Selection of Papers from the Review of Economic Studies. Palgrave Macmillan UK, London, pp. 131–149. https://doi.org/10.1007/978-1-349-15430-2_11

Audretsch, D.B., Feldman, M.P., 2004. Chapter 61 - Knowledge Spillovers and the Geography of Innovation, in: Henderson, J.V., Thisse, J.-F. (Eds.), Handbook of Regional and Urban Economics, Cities and Geography. Elsevier, pp. 2713–2739. https://doi.org/10.1016/S1574-0080(04)80018-X

Audretsch, D.B., Feldman, M.P., 1996. R&D Spillovers and the Geography of Innovation and Production. The American Economic Review 86, 630–640.

Bahar, D., Choudhury, P., Rapoport, H., 2020. Migrant inventors and the technological advantage of nations. Research Policy, STEM migration, research, and innovation 49, 103947. https://doi.org/10.1016/j.respol.2020.103947

Bahar, D., Hausmann, R., Hidalgo, C.A., 2014. Neighbors and the evolution of the comparative advantage of nations: Evidence of international knowledge diffusion? Journal of International Economics 92, 111–123. https://doi.org/10.1016/j.jinteco.2013.11.001

Bahar, D., Rosenow, S., Stein, E., Wagner, R., 2019. Export take-offs and acceleration: Unpacking cross-sector linkages in the evolution of comparative advantage. World Development 117, 48–60.

Balassa, B., 1985. Exports, policy choices, and economic growth in developing countries after the 1973 oil shock. Journal of Development Economics 18, 23–35. https://doi.org/10.1016/0304-3878(85)90004-5

Balassa, B., 1978. Exports and economic growth: Further evidence. Journal of Development Economics 5, 181–189. https://doi.org/10.1016/0304-3878(78)90006-8

Balland, P.A., Boschma, R., Crespo, J., Rigby, D.L., 2018. Smart specialization policy in the European Union: relatedness, knowledge complexity and regional diversification. Regional Studies 1–17. https://doi.org/10.1080/00343404.2018.1437900

Balland, P.-A., Jara-Figueroa, C., Petralia, S.G., Steijn, M.P.A., Rigby, D.L., Hidalgo, C.A., 2020. Complex economic activities concentrate in large cities. Nat Hum Behav 1–7. https://doi.org/10.1038/s41562-019-0803-3

Balland, P.-A., Rigby, D., 2017. The Geography of Complex Knowledge. Economic Geography 93, 1–23. https://doi.org/10.1080/00130095.2016.1205947

Bandeira Morais, M., Swart, J., Jordaan, J.A., 2018. Economic Complexity and Inequality: Does Productive Structure Affect Regional Wage Differentials in Brazil? USE Working Paper series 18.

Bank, A.D., 2017. Asian Development Outlook (ADO) 2017: Transcending the Middle-Income Challenge. Asian Development Bank. http://dx.doi.org/10.22617/FLS178632-3

Barza, R., Jara-Figueroa, C., Hidalgo, C., Viarengo, M., 2020. Knowledge Intensity and Gender Wage Gaps: Evidence from Linked Employer-Employee Data.

Basile, R., Cicerone, G., 2022. Economic complexity and productivity polarization: Evidence from Italian provinces. German Economic Review. https://doi.org/10.1515/ger-2021-0070

Basile, R., Cicerone, G., Iapadre, L., 2019. Economic complexity and regional labor productivity distribution: evidence from Italy.

Beine, M., Docquier, F., Rapoport, H., 2008. Brain drain and human capital formation in developing countries: winners and losers. The Economic Journal 118, 631–652.

Bell, M., Pavitt, K., 1993. Technological accumulation and industrial growth: Contrasts between.

Ben Saâd, M., Assoumou-Ella, G., 2019. Economic Complexity and Gender Inequality in Education: An Empirical Study. Economics Bulletin 39, 321–334.

Berger, T., Enflo, K., 2017. Locomotives of local growth: The short- and long-term impact of railroads in Sweden. Journal of Urban Economics, Urbanization in Developing Countries: Past and Present 98, 124–138. https://doi.org/10.1016/j.jue.2015.09.001

Bishop, A., Mateos-Garcia, J., 2019. Exploring the Link Between Economic Complexity and Emergent Economic Activities. National Institute Economic Review 249, R47–R58.

Bontadini, F., Savona, M., 2017. Revisiting the Natural Resource Industries "curse": Beneficiation or Hirschman Backward Linkages?

Boschma, R., 2005. Proximity and Innovation: A Critical Assessment. Regional Studies 39, 61–74. https://doi.org/10.1080/0034340052000320887

Boschma, R., Balland, P.-A., Kogler, D.F., 2015. Relatedness and technological change in cities: the rise and fall of technological knowledge in US metropolitan areas from 1981 to 2010. Ind Corp Change 24, 223–250. https://doi.org/10.1093/icc/dtu012

Boschma, R., Capone, G., 2015. Institutions and diversification: Related versus unrelated diversification in a varieties of capitalism framework. Research Policy 44, 1902–1914. https://doi.org/10.1016/j.respol.2015.06.013

Boschma, R., Frenken, K., 2011. The emerging empirics of evolutionary economic geography. J Econ Geogr 11, 295–307. https://doi.org/10.1093/jeg/lbq053

Boschma, R., Frenken, K., Bathelt, H., Feldman, M., Kogler, D., 2012. Technological relatedness and regional branching. Beyond territory. Dynamic geographies of knowledge creation, diffusion and innovation 64–68.

Boschma, R., Minondo, A., Navarro, M., 2013. The Emergence of New Industries at the Regional Level in Spain: A Proximity Approach Based on Product Relatedness. Economic Geography 89, 29–51. https://doi.org/10.1111/j.1944-8287.2012.01170.x

Breschi, S., Lissoni, F., 2004. Knowledge networks from patent data, in: Handbook of Quantitative Science and Technology Research. Springer, pp. 613–643.

Breschi, S., Lissoni, F., Miguelez, E., 2018. Return migrants' self-selection: Evidence for Indian inventor. National Bureau of Economic Research.

Breschi, S., Lissoni, F., Miguelez, E., 2017. Foreign-origin inventors in the USA: testing for diaspora and brain gain effects. Journal of Economic Geography 17, 1009–1038.

Britto, G., ROMERO, J., FREITAS, E., COELHO, C., 2016. The Great Divide: Economic Complexity and Development Paths in Brazil and South Korea. Blucher Engineering Proceedings 3, 1404–1425.

Brücker, H., Docquier, F., Rapoport, H., 2012. Brain Drain and Brain Gain: The Global Competition to Attract High-Skilled Migrants. OUP Oxford.

Bruno, M., Saracco, F., Squartini, T., Dueñas, M., 2018. Colombian export capabilities: building the firms-products network. Entropy 20, 785.

Brynjolfsson, E., Hui, X., Liu, M., 2019. Does machine translation affect international trade? Evidence from a large digital platform. Management Science 65, 5449–5460.

Bustos, S., Gomez, C., Hausmann, R., Hidalgo, C.A., 2012. The Dynamics of Nestedness Predicts the Evolution of Industrial Ecosystems. PLOS ONE 7, e49393. https://doi.org/10.1371/journal.pone.0049393

Bustos, S., Yıldırım, M.A., 2022. Production ability and economic growth. Research Policy 51, 104153.

Can, M., Gozgor, G., 2017. The impact of economic complexity on carbon emissions: evidence from France. Environmental Science and Pollution Research 24, 16364–16370.

Catalán, P., Navarrete, C., Figueroa, F., 2020. The scientific and technological cross-space: Is technological diversification driven by scientific endogenous capacity? Research Policy 104016. https://doi.org/10.1016/j.respol.2020.104016

Catalini, C., Fons-Rosen, C., Gaulé, P., 2020. How Do Travel Costs Shape Collaboration? Management Science 66, 3340–3360. https://doi.org/10.1287/mnsc.2019.3381

Caviggioli, F., Jensen, P., Scellato, G., 2020. Highly skilled migrants and technological diversification in the US and Europe. Technological Forecasting and Social Change 154, 119951. https://doi.org/10.1016/j.techfore.2020.119951

Chang, H.-J., 1994. State, institutions and structural change. Structural Change and Economic Dynamics 5, 293–313.

Chávez, J.C., Mosqueda, M.T., Gómez-Zaldívar, M., 2017. Economic complexity and regional growth performance: Evidence from the Mexican Economy. Review of Regional Studies 47, 201–219.

Chen, Z., Poncet, S., Xiong, R., 2017. Inter-industry relatedness and industrial-policy efficiency: Evidence from China's export processing zones. Journal of Comparative Economics 45, 809–826. https://doi.org/10.1016/j.jce.2016.01.003

Choudhury, P., 2016. Return migration and geography of innovation in MNEs: a natural experiment of knowledge production by local workers reporting to return migrants. Journal of Economic Geography 16, 585–610. https://doi.org/10.1093/jeg/lbv025

Chu, L.K., Hoang, D.P., 2020. How does economic complexity influence income inequality? New evidence from international data. Economic Analysis and Policy 68, 44–57.

Cicerone, G., McCann, P., Venhorst, V.A., 2020. Promoting regional growth and innovation: relatedness, revealed comparative advantage and the product space. Journal of Economic Geography 20, 293–316.

Cohen, W.M., Levinthal, D.A., 1990. Absorptive Capacity: A New Perspective on Learning and Innovation. Administrative Science Quarterly 35, 128–152. https://doi.org/10.2307/2393553

Coscia, M., Neffke, F.M., Hausmann, R., 2020. Knowledge diffusion in the network of international business travel. Nature Human Behaviour 4, 1011–1020.

Coskun, N., Lopcu, K., Tuncer, İ., 2018. The Economic Complexity Approach to Development Policy: Where Turkey Stands in Comparison to OECD plus China? Proceedings of Middle East Economic Association 20, 112–124.

Crescenzi, R., Dyevre, A., Neffke, F., 2020. Innovation catalysts: how multinationals reshape the global geography of innovation (No. 105684), LSE Research Online Documents on Economics, LSE Research Online Documents on Economics. London School of Economics and Political Science, LSE Library.

Dasgupta, P., Stiglitz, J., 1988. Learning-by-doing, market structure and industrial and trade policies. Oxford Economic Papers 40, 246–268.

DataMéxico [WWW Document], 2020. . Data México. URL https://datamexico.org/ (accessed 10.5.21).

De Waldemar, F.S., Poncet, S., 2013. Product relatedness and firm exports in China. The world bank economic review 51, 104Ð118.

Deegan, J., Broekel, T., Fitjar, R.D., 2021. Searching through the Haystack:The Relatedness and Complexity of Priorities in Smart Specialization Strategies. Economic Geography 97, 497–520. https://doi.org/10.1080/00130095.2021.1967739

Diodato, D., Neffke, F., O'Clery, N., 2018. Why do industries coagglomerate? How Marshallian externalities differ by industry and have evolved over time. Journal of Urban Economics 106, 1–26.

Dittmar, J.E., 2011. Information technology and economic change: the impact of the printing press. The Quarterly Journal of Economics 126, 1133–1172.

Doğan, B., Ghosh, S., Shahzadi, I., Balsalobre-Lorente, D., Nguyen, C.P., 2022. The relevance of economic complexity and economic globalization as determinants of energy demand for different stages of development. Renewable Energy. https://doi.org/10.1016/j.renene.2022.03.117

Domini, G., 2019. Patterns of specialisation and economic complexity through the lens of universal exhibitions, 1855-1900. LEM Working Paper Series.

Dong, Z., Chen, W., Wang, S., 2020. Emission reduction target, complexity and industrial performance. Journal of Environmental Management 260, 110148.

Dong, Z., Li, Y., Balland, P.-A., Zheng, S., 2022. Industrial land policy and economic complexity of Chinese Cities. Industry and Innovation 29, 367–395.

Dordmond, G., de Oliveira, H.C., Silva, I.R., Swart, J., 2020. The complexity of Green job creation: An Analysis of green job development in Brazil. Environment, Development and Sustainability 1–24.

Economía, S. de, 2021. Diversificación inteligente [WWW Document]. gob.mx. URL http://www.gob.mx/se/acciones-y-programas/diversificacion-inteligente (accessed 4.14.22).

Eisenstein, E.L., 1980. The Printing Press as an Agent of Change: Communications and Cultural Trans. Cambridge University Press, Cambridge.

Elekes, Z., Boschma, R., Lengyel, B., 2019. Foreign-owned firms as agents of structural change in regions. Regional Studies 53, 1603–1613. https://doi.org/10.1080/00343404.2019.1596254

Ellison, G., Glaeser, E.L., Kerr, W.R., 2010. What Causes Industry Agglomeration? Evidence from Coagglomeration Patterns. American Economic Review 100, 1195–1213. https://doi.org/10.1257/aer.100.3.1195

Erkan, B., Yildirimci, E., 2015. Economic Complexity and Export Competitiveness: The Case of Turkey. Procedia-Social and Behavioral Sciences 195, 524–533.

Essletzbichler, J., 2015. Relatedness, industrial branching and technological cohesion in US metropolitan areas. Regional Studies 49, 752–766.

Farinha, T., Balland, P.-A., Morrison, A., Boschma, R., 2019. What drives the geography of jobs in the us? unpacking relatedness. Industry and Innovation 26, 988–1022.

Fawaz, F., Rahnama-Moghadamm, M., 2019. Spatial dependence of global income inequality: The role of economic complexity. The International Trade Journal 33, 542–554.

Felipe, J., Abdon, A., Kumar, U., 2012a. Tracking the Middle-Income Trap: What is it, Who is in it, and Why? (SSRN Scholarly Paper No. ID 2049330). Social Science Research Network, Rochester, NY. https://doi.org/10.2139/ssrn.2049330

Felipe, J., Kumar, U., Abdon, A., Bacate, M., 2012b. Product complexity and economic development. Structural Change and Economic Dynamics 23, 36–68. https://doi.org/10.1016/j.strueco.2011.08.003

Fernandes, A.M., Mattoo, A., Nguyen, H., Schiffbauer, M., 2019. The internet and Chinese exports in the pre-ali baba era. Journal of Development Economics 138, 57–76.

Ferrarini, B., Scaramozzino, P., 2015. The product space revisited: China's trade profile. The World Economy 38, 1368–1386.

Ferreira-Coimbra, N., Vaillant, M., 2009. Evolución del espacio de productos exportados:¿ está Uruguay en el lugar equivocado? Revista de economía 16, 97–146.

Foray, D., David, P.A., Hall, B., 2009. Smart specialisation–the concept. Knowledge economists policy brief 9, 100.

Fraccascia, L., Giannoccaro, I., Albino, V., 2018. Green product development: What does the country product space imply? Journal of cleaner production 170, 1076–1088.

Freeman, C., 2002. Continental, national and sub-national innovation systems-complementarity and economic growth. Research Policy 31, 191–191.

Fritz, B.S., Manduca, R.A., 2021. The economic complexity of US metropolitan areas. Regional Studies 1–12.

Fukuyama, F., 1989. The end of history? The national interest 3–18.

Gala, P., 2017. Complexidade Economica: Uma Nova Perspectiva Para Entender a Antiga Questao da Riqueza das Nacoes. Contraponto.

Gao, J., Jun, B., Pentland, A. 'Sandy,' Zhou, T., Hidalgo, C.A., 2021. Spillovers across industries and regions in China's regional economic diversification. Regional Studies 1–16.

Gao, J., Zhou, T., 2018. Quantifying China's regional economic complexity. Physica A: Statistical Mechanics and its Applications 492, 1591–1603. https://doi.org/10.1016/j.physa.2017.11.084

Gerschenkron, A., 2015. Economic backwardness in historical perspective (1962). Cambridge MA.

Gerschenkron, A., 1963. The early phases of industrialization in Russia: afterthoughts and counterthoughts, in: The Economics of Take-off into Sustained Growth. Springer, pp. 151–169.

Gmelch, G., 1980. Return migration. Annual review of anthropology 135–159.

Gnecco, G., Nutarelli, F., Riccaboni, M., 2022. A machine learning approach to economic complexity based on matrix completion. Scientific Reports 12, 9639.

González, A., Ortigoza, E., Llamosas, C., Blanco, G., Amarilla, R., 2018. Multi-criteria analysis of economic complexity transition in emerging economies: The case of Paraguay. Socio-Economic Planning Sciences 100617.

Guevara, M.R., Hartmann, D., Aristarán, M., Mendoza, M., Hidalgo, C.A., 2016. The research space: using career paths to predict the evolution of the research output of individuals, institutions, and nations. Scientometrics 109, 1695–1709. https://doi.org/10.1007/s11192-016-2125-9

Guo, Q., He, C., 2017. Production space and regional industrial evolution in China. GeoJournal 82, 379–396. https://doi.org/10.1007/s10708-015-9689-4

Hamilton, A., 1791. Report on manufactures.

Hamwey, R., Pacini, H., Assunção, L., 2013. Mapping green product spaces of nations. The Journal of Environment & Development 22, 155–168.

Hartmann, D., 2016. The economic diversification and innovation system of Turkey from a global comparative perspective, in: International Innovation Networks and Knowledge Migration. Routledge, pp. 53–71.

Hartmann, D., Guevara, M.R., Jara-Figueroa, C., Aristarán, M., Hidalgo, C.A., 2017. Linking Economic Complexity, Institutions, and Income Inequality. World Development 93, 75–93. https://doi.org/10.1016/j.worlddev.2016.12.020

Harvey, D.I., Kellard, N.M., Madsen, J.B., Wohar, M.E., 2010. The Prebisch-Singer hypothesis: four centuries of evidence. The review of Economics and Statistics 92, 367–377.

Hassink, R., Gong, H., 2019. Six critical questions about smart specialization. European Planning Studies 27, 2049–2065. https://doi.org/10.1080/09654313.2019.1650898

Hausmann, R., Hidalgo, C.A., 2011. The network structure of economic output. Journal of Economic Growth 1–34.

Hausmann, R., Hidalgo, C.A., Bustos, S., Coscia, M., Simoes, A., Yildirim, M.A., 2014. The atlas of economic complexity: Mapping paths to prosperity. MIT Press.

Hausmann, R., Hidalgo, C.A., Jiménez, J., Lawrence, R., Yeyati, E.L., Sabel, C., Schydlowsky, D., 2011. Construyendo un mejor futuro para la República Dominicana: herramientas para el desarrollo. Informe técnico. Cambridge, MA: Center for International Development, Universidad de Harvard.

Hausmann, R., Hwang, J., Rodrik, D., 2007. What you export matters. J Econ Growth 12, 1–25. https://doi.org/10.1007/s10887-006-9009-4

Hausmann, R., Stock, D.P., Yıldırım, M.A., 2021. Implied comparative advantage. Research Policy 104143.

He, C., Zhu, S., 2019. Evolutionary Economic Geography in China, Economic Geography. Springer Singapore.

Hidalgo, Balland, P.-A., Boschma, R., Delgado, M., Feldman, M., Frenken, K., Glaeser, E., He, C., Kogler, D.F., Morrison, A., Neffke, F., Rigby, D., Stern, S., Zheng, S., Zhu, S., 2018. The Principle of Relatedness, in: Morales, A.J., Gershenson, C., Braha, D., Minai, A.A., Bar-Yam, Y. (Eds.), Unifying Themes in Complex Systems IX, Springer Proceedings in Complexity. Springer International Publishing, pp. 451–457.

Hidalgo, C., 2015. Why information grows: The evolution of order, from atoms to economies. Basic Books, New York.

Hidalgo, C.A., 2022. Knowledge is non-fungible. arXiv preprint arXiv:2205.02167.

Hidalgo, C.A., 2021. Economic complexity theory and applications. Nature Reviews Physics 1–22.

Hidalgo, C.A., Castañer, E., Sevtsuk, A., 2020. The amenity mix of urban neighborhoods. Habitat International 102205. https://doi.org/10.1016/j.habitatint.2020.102205

Hidalgo, C.A., Hausmann, R., 2009. The building blocks of economic complexity. PNAS 106, 10570–10575. https://doi.org/10.1073/pnas.0900943106

Hidalgo, C.A., Klinger, B., Barabási, A.-L., Hausmann, R., 2007. The Product Space Conditions the Development of Nations. Science 317, 482–487. https://doi.org/10.1126/science.1144581

Hirschman, A.O., 1977. A generalized linkage approach to development, with special reference to staples. Economic development and cultural change 25, 67.

Hjort, J., Poulsen, J., 2019. The arrival of fast internet and employment in Africa. American Economic Review 109, 1032–79.

Hovhannisyan, N., Keller, W., 2015. International business travel: an engine of innovation? J Econ Growth 20, 75–104. https://doi.org/10.1007/s10887-014-9107-7

Imbs, J., Wacziarg, R., 2003. Stages of diversification. American economic review 93, 63–86.

Innis, H.A., 2008. The bias of communication. University of Toronto Press.

Innocenti, N., Lazzeretti, L., 2019a. Do the creative industries support growth and innovation in the wider economy? Industry relatedness and employment growth in Italy. Industry and Innovation 26, 1152–1173.

Innocenti, N., Lazzeretti, L., 2019b. Growth in regions, knowledge bases and relatedness: some insights from the Italian case. European Planning Studies 27, 2034–2048.

Innocenti, N., Lazzeretti, L., 2017. Related variety and employment growth in Italy. Scienze Regionali 16, 325–350.

Irwin, D.A., 2021. The rise and fall of import substitution. World Development 139, 105306.

Jaffe, A.B., 1989. Real effects of academic research. American economic review 79, 957–970.

Jaffe, A.B., 1986. Technological opportunity and spillovers of R&D: evidence from firms' patents, profits and market value. national bureau of economic research Cambridge, Mass., USA.

Jaffe, A.B., Trajtenberg, M., Henderson, R., 1993. Geographic Localization of Knowledge Spillovers as Evidenced by Patent Citations. Q J Econ 108, 577–598. https://doi.org/10.2307/2118401

Jara-Figueroa, C., Jun, B., Glaeser, E.L., Hidalgo, C.A., 2018. The role of industry-specific, occupation-specific, and location-specific knowledge in the growth and survival of new firms. PNAS 115, 12646–12653. https://doi.org/10.1073/pnas.1800475115

Jara-Figueroa, C., Yu, A.Z., Hidalgo, C.A., 2019. How the medium shapes the message: Printing and the rise of the arts and sciences. PLOS ONE 14, e0205771. https://doi.org/10.1371/journal.pone.0205771

Jun, B., Alshamsi, A., Gao, J., Hidalgo, C.A., 2019. Bilateral relatedness: knowledge diffusion and the evolution of bilateral trade. Journal of Evolutionary Economics 1–31.

Kahn, M.E., Sun, W., Wu, J., Zheng, S., 2018. The Revealed Preference of the Chinese Communist Party Leadership: Investing in Local Economic Development versus Rewarding Social Connections (Working Paper No. 24457). National Bureau of Economic Research. https://doi.org/10.3386/w24457

Karsten, Jack, 2022. Building a Stronger (More Complex) U.S. Manufacturing Sector. Innovation Frontier Project. URL https://innovationfrontier.org/building-a-stronger-more-complex-u-s-manufacturing-sector/ (accessed 4.14.22).

Kerr, W.R., 2018. The Gift of Global Talent: How Migration Shapes Business, Economy & Society. Stanford University Press.

Kerr, W.R., 2008. Ethnic scientific communities and international technology diffusion. The Review of Economics and Statistics 90, 518–537.

Kleinberg, J., Ludwig, J., Mullainathan, S., Obermeyer, Z., 2015. Prediction policy problems. American Economic Review 105, 491–95.

Koch, P., 2021. Economic Complexity and Growth: Can value-added exports better explain the link? Economics Letters 198, 109682.

Kogler, D.F., Rigby, D.L., Tucker, I., 2013. Mapping Knowledge Space and Technological Relatedness in US Cities. European Planning Studies 21, 1374–1391. https://doi.org/10.1080/09654313.2012.755832

Kojima, K., 2000. The "flying geese" model of Asian economic development: origin, theoretical extensions, and regional policy implications. Journal of Asian Economics 11, 375–401.

Krugman, P.R., 1993. What Do Undergrads Need to Know About Trade? The American Economic Review 83, 23–26.

Lapatinas, A., 2019. The effect of the Internet on economic sophistication: An empirical analysis. Economics Letters 174, 35–38.

Lapatinas, A., Garas, A., Boleti, E., Kyriakou, A., 2019. Economic complexity and environmental performance: Evidence from a world sample.

Lee, K., Lim, C., 2001. Technological regimes, catching-up and leapfrogging: findings from the Korean industries. Research Policy 30, 459–483. https://doi.org/10.1016/S0048-7333(00)00088-3

Lee, K., Malerba, F., 2017. Catch-up cycles and changes in industrial leadership:Windows of opportunity and responses of firms and countries in the evolution of sectoral systems. Research Policy 46, 338–351. https://doi.org/10.1016/j.respol.2016.09.006

Lin, J., Chang, H.-J., 2009. Should Industrial Policy in developing countries conform to comparative advantage or defy it? A debate between Justin Lin and Ha-Joon Chang. Development policy review 27, 483–502.

Lin, J.Y., 2011. New structural economics: A framework for rethinking development. The World Bank Research Observer 26, 193–221.

Lo Turco, A., Maggioni, D., 2020. The knowledge and skill content of production complexity. Research Policy 104059. https://doi.org/10.1016/j.respol.2020.104059

Lo Turco, A., Maggioni, D., 2019. Local discoveries and technological relatedness: the role of MNEs, imports and domestic capabilities. Journal of Economic Geography 19, 1077–1098. https://doi.org/10.1093/jeg/lby060

López González, J., Meliciani, V., Savona, M., 2019. When Linder meets Hirschman: inter-industry linkages and global value chains in business services. Industrial and Corporate Change 28, 1555–1586. https://doi.org/10.1093/icc/dtz023

Lybbert, T.J., Xu, M., 2022. Innovation-adjusted economic complexity and growth: Do patent flows reveal enhanced economic capabilities? Review of Development Economics 26, 442–483. https://doi.org/10.1111/rode.12816

Lyubimov, I., Gvozdeva, M., Kazakova, M., Nesterova, K., 2017. Economic Complexity of Russian Regions and their Potential to Diversify. Journal of the New Economic Association 34, 94–122.

Lyubimov, I.L., Lysyuk, M.V., Gvozdeva, M.A., 2018. Atlas of economic complexity, Russian regional pages. VOPROSY ECONOMIKI 6.

MacKinnon, D., Dawley, S., Pike, A., Cumbers, A., 2019. Rethinking path creation: A geographical political economy approach. Economic Geography 95, 113–135.

Maes, P., 1995. Agents that Reduce work and information Overload, in: Baecker, R.M., Grudin, J., Buxton, W.A.S., Greenberg, S. (Eds.), Readings in Human–Computer Interaction, Interactive Technologies. Morgan Kaufmann, pp. 811–821. https://doi.org/10.1016/B978-0-08-051574-8.50084-4

Malerba, F., Lee, K., 2021. An evolutionary perspective on economic catch-up by latecomers. Industrial and Corporate Change 30, 986–1010.

Mayr, K., Peri, G., 2008. Return migration as a channel of brain gain. National Bureau of Economic Research.

Mazzucato, M., 2018. Mission-oriented innovation policies: challenges and opportunities. Industrial and Corporate Change 27, 803–815. https://doi.org/10.1093/icc/dty034

McNerney, J., Li, Y., Gomez-Lievano, A., Neffke, F., 2023. Bridging the short-term and long-term dynamics of economic structural change. https://doi.org/10.48550/arXiv.2110.09673

Mealy, P., Coyle, D., 2021. To them that hath: economic complexity and local industrial strategy in the UK. International Tax and Public Finance 1–20.

Mealy, P., Farmer, J.D., Teytelboym, A., 2019. Interpreting economic complexity. Science Advances 5, eaau1705. https://doi.org/10.1126/sciadv.aau1705

Mealy, P., Teytelboym, A., 2020. Economic complexity and the green economy. Research Policy 103948.

Miguelez, E., Moreno, R., 2018. Relatedness, external linkages and regional innovation in Europe. Regional Studies 52, 688–701. https://doi.org/10.1080/00343404.2017.1360478

Miguelez, E., Morrison, A., 2022. Migrant inventors as agents of technological change. The Journal of Technology Transfer 1–24.

Miguelez, E., Noumedem Temgoua, C., 2020. Inventor migration and knowledge flows: A two-way communication channel? Research Policy, STEM migration, research, and innovation 49, 103914. https://doi.org/10.1016/j.respol.2019.103914

Montresor, S., Quatraro, F., 2019. Green technologies and Smart Specialisation Strategies: a European patent-based analysis of the intertwining of technological relatedness and key enabling technologies. Regional Studies 1–12.

Moreno, R., Ocampo-Corrales, D., 2022. The ability of European regions to diversify in renewable energies: The role of technological relatedness. Research Policy 51, 104508.

Mu, Q., Lee, K., 2005. Knowledge diffusion, market segmentation and technological catch-up: The case of the telecommunication industry in China. Research policy 34, 759–783.

Neagu, O., 2019. The Link between Economic Complexity and Carbon Emissions in the European Union Countries: A Model Based on the Environmental Kuznets Curve (EKC) Approach. Sustainability 11, 4753.

Neffke, F., Hartog, M., Boschma, R., Henning, M., 2018. Agents of Structural Change: The Role of Firms and Entrepreneurs in Regional Diversification. Economic Geography 94, 23–48. https://doi.org/10.1080/00130095.2017.1391691

Neffke, F., Henning, M., 2013. Skill relatedness and firm diversification. Strategic Management Journal 34, 297–316.

Neffke, F., Henning, M., Boschma, R., 2011. How Do Regions Diversify over Time? Industry Relatedness and the Development of New Growth Paths in Regions. Economic Geography 87, 237–265.

Nelson, R.R., Dosi, G., Helfat, C.E., Pyka, A., Saviotti, P.P., Lee, K., Dopfer, K., Malerba, F., Winter, S.G., 2018. Modern Evolutionary Economics: An Overview. Cambridge University Press, Cambridge, United Kingdom ; New York, NY.

Nelson, R.R., Nelson, K., 2002. Technology, institutions, and innovation systems. Research policy 31, 265–272.

Nelson, R.R., Winter, S.G., 1985. An Evolutionary Theory of Economic Change. Belknap Press: An Imprint of Harvard University Press, Cambridge, Mass.

Nereim, Vivian, 2022. MBS's $500 Billion Desert Dream Just Keeps Getting Weirder. Bloomberg.com.

Ning, L., Guo, R., 2022. Technological diversification to green domains: technological relatedness, invention impact and knowledge integration capabilities. Research Policy 51, 104406.

Ourens, G., 2012. Can the Method of Reflections help predict future growth? Documento de Trabajo/FCS-DE; 17/12.

Parsons, C., Vézina, P.-L., 2017. Migrant Networks and Trade: The Vietnamese Boat People as a Natural Experiment. Economic Journal.

Pérez Hernández, C.C., Salazar Hernández, B.C., Mendoza Moheno, J., 2019. Diagnóstico de la complejidad económica del estado de Hidalgo: de las capacidades a las oportunidades. Revista mexicana de economía y finanzas 14, 261–277.

Perruchas, F., Consoli, D., Barbieri, N., 2020. Specialisation, diversification and the ladder of green technology development. Research Policy 49, 103922.

Pettegree, A., 2010. The book in the Renaissance. JSTOR.

Pinheiro, F.L., Hartmann, D., Boschma, R., Hidalgo, C.A., 2021. The time and frequency of unrelated diversification. Research Policy 104323.

Poncet, S., de Waldemar, F.S., 2013. Economic Complexity and Growth. Revue économique 64, 495–503.

Porter, M.E., 1998. Clusters and the new economics of competition. Harvard Business Review.

Porter, M.E., 1990. The Competitive Advantage of Nations. Simon and Shuster.

Prebisch, R., 1962. The economic development of Latin America and its principal problems. Economic Bulletin for Latin America.

Pugliese, E., Cimini, G., Patelli, A., Zaccaria, A., Pietronero, L., Gabrielli, A., 2019a. Unfolding the innovation system for the development of countries: coevolution of Science, Technology and Production. Scientific reports 9, 1–12.

Pugliese, E., Napolitano, L., Zaccaria, A., Pietronero, L., 2019b. Coherent diversification in corporate technological portfolios. PloS one 14, e0223403.

Pugliese, E., Tacchella, A., 2020. Economic complexity for competitiveness and innovation: A novel bottom-up strategy linking global and regional capacities. Joint Research Centre (Seville site).

Rapoport, H., 2016. Migration and globalization: what's in it for developing countries? Int J of Manpower 37, 1209–1226. https://doi.org/10.1108/IJM-08-2015-0116

Rapoport, H., 2004. Who is afraid of the brain drain? Human capital flight and growth in developing countries. Brussels Economic Review 47, 89–101.

Resnick, P., Varian, H.R., 1997. Recommender systems. Communications of the ACM 40, 56–58.

Reynolds, C., Agrawal, M., Lee, I., Zhan, C., Li, J., Taylor, P., Mares, T., Morison, J., Angelakis, N., Roos, G., 2018. A sub-national economic complexity analysis of Australia's states and territories. Regional Studies 52, 715–726.

Rigby, D.L., 2015. Technological relatedness and knowledge space: Entry and exit of US cities from patent classes. Regional Studies 49, 1922–1937.

Rodrik, D., 2006. What's so special about China's exports? China & World Economy 14, 1–19.

Romer, P.M., 1990. Endogenous Technological Change. Journal of Political Economy 98, S71–S102. https://doi.org/10.1086/261725

Romer, P.M., 1986. Increasing Returns and Long-Run Growth. Journal of Political Economy 94, 1002–1037. https://doi.org/10.1086/261420

Romero, J.P., Gramkow, C., 2021. Economic complexity and greenhouse gas emissions. World Development 139, 105317. https://doi.org/10.1016/j.worlddev.2020.105317

Rosenstein-Rodan, P.N., 1961. Notes on the theory of the 'big push,' in: Economic Development for Latin America. Springer, pp. 57–81.

Rosenstein-Rodan, P.N., 1943. Problems of industrialisation of eastern and south-eastern Europe. The economic journal 53, 202–211.

Rostow, W.W., 1959. The stages of economic growth. The economic history review 12, 1–16.

Santoalha, A., Boschma, R., 2020. Diversifying in green technologies in European regions: does political support matter? Regional Studies 1–14.

Saviotti, P.P., Frenken, K., 2008. Export variety and the economic performance of countries. Journal of Evolutionary Economics 18, 201–218.

Saviotti, P.P., Pyka, A., 2004. Economic development by the creation of new sectors. Journal of evolutionary economics 14, 1–35.

Savona, M., 2018. Industrial policy for a European industrial renaissance. A few reflections. A Few Reflections (March 6, 2018). SWPS 7.
Saxenian, A., 2007. The New Argonauts: Regional Advantage in a Global Economy. Harvard University Press.
Sbardella, A., Nicolò, B., Davide, C., Lorenzo, N., François, P., Emanuele, P., 2022. The regional green potential of the European innovation system. Joint Research Centre (Seville site).
Sbardella, A., Pugliese, E., Pietronero, L., 2017. Economic development and wage inequality: A complex system analysis. PloS one 12.
Sciarra, C., Chiarotti, G., Ridolfi, L., Laio, F., 2020. Reconciling contrasting views on economic complexity. Nature communications 11, 1–10.
Singer, H.W., 2012. The distribution of gains between investing and borrowing countries, in: Milestones and Turning Points in Development Thinking. Springer, pp. 265–277.
Singh, J., 2005. Collaborative Networks as Determinants of Knowledge Diffusion Patterns. Management Science 51, 756–770. https://doi.org/10.1287/mnsc.1040.0349
Solow, R.M., 1956. A Contribution to the Theory of Economic Growth. Q J Econ 70, 65–94. https://doi.org/10.2307/1884513
Stafforte, S., Tamberi, M., 2012. Italy in the space (of products). Economia Marche Journal of Applied Economics 31, 90–113.
Stiglitz, J.E., 2017. Industrial policy, learning and development.
Stojkoski, V., Kocarev, L., 2017. The relationship between growth and economic complexity: evidence from Southeastern and Central Europe.
Stojkoski, V., Koch, P., Hidalgo, C.A., 2022. Multidimensional Economic Complexity: How the Geography of Trade, Technology, and Research Explain Inclusive Green Growth. arXiv preprint arXiv:2209.08382.
Stojkoski, V., Utkovski, Z., Kocarev, L., 2016. The Impact of Services on Economic Complexity: Service Sophistication as Route for Economic Growth. PLOS ONE 11, e0161633. https://doi.org/10.1371/journal.pone.0161633
Swart, J., Brinkmann, L., 2020. Economic Complexity and the Environment: Evidence from Brazil, in: Universities and Sustainable Communities: Meeting the Goals of the Agenda 2030. Springer, pp. 3–45.
Tacchella, A., Cristelli, M., Caldarelli, G., Gabrielli, A., Pietronero, L., 2012. A New Metrics for Countries' Fitness and Products' Complexity. Scientific Reports 2, srep00723. https://doi.org/10.1038/srep00723
Tacchella, A., Zaccaria, A., Miccheli, M., Pietronero, L., 2021. Relatedness in the era of machine learning. arXiv preprint arXiv:2103.06017.
Teece, D.J., Rumelt, R., Dosi, G., Winter, S., 1994. Understanding corporate coherence: Theory and evidence. Journal of economic behavior & organization 23, 1–30.
Torre, A., Rallet, A., 2005. Proximity and localization. Regional studies 39, 47–59.
Tóth, G., Lengyel, B., 2021. Inter-firm inventor mobility and the role of co-inventor networks in producing high-impact innovation. J Technol Transf 46, 117–137. https://doi.org/10.1007/s10961-019-09758-5
Tullio, B., Giancarlo, C., 2020. Foreign Direct Investment and Economic Complexity: Key Elements for a Targeted Industrial Policy in Italy. L'industria 769–786.

Uhlbach, W.-H., Balland, P.-A., Scherngell, T., 2017. R&D Policy and Technological Trajectories of Regions: Evidence from the EU Framework Programmes.

Uyarra, E., Flanagan, K., 2021. Going beyond the line of sight: institutional entrepreneurship and system agency in regional path creation. Regional Studies 1–12.

van der Wouden, F., Rigby, D.L., 2019. Co-inventor networks and knowledge production in specialized and diversified cities. Papers in Regional Science 98, 1833–1853.

Vega-Villa, K.R., 2017. Missed opportunities in Yachay. Science 358, 459–459.

Wang, Y., Turkina, E., 2020a. Economic Complexity and Industrial Upgrading in the Product Space Network-Opportunities for the City of Laval, Canada, in: Rethinking Cluster 2020 Conference.

Wang, Y., Turkina, E., 2020b. Economic complexity, product space network and Quebec's global competitiveness. Canadian Journal of Administrative Sciences/Revue Canadienne des Sciences de l'Administration 37, 334–349.

Waniek, M., Elbassioni, K., Pinheiro, F.L., Hidalgo, C.A., Alshamsi, A., 2020. Computational aspects of optimal strategic network diffusion. Theoretical Computer Science.

Weitzman, M.L., 1998. Recombinant Growth. Q J Econ 113, 331–360. https://doi.org/10.1162/003355398555595

Williamson, J., 2009. A short history of the Washington Consensus. Law & Bus. Rev. Am. 15, 7.

Zaldívar, F.G., Molina, E., Flores, M., Zaldívar, M. de J.G., 2019. Economic Complexity of the Special Economic Zones in Mexico: Opportunities for Diversification and Industrial Sophistication. Ensayos Revista de Economía (Ensayos Journal of Economics) 38, 1–40.

Zheng, S., Sun, W., Wu, J., Kahn, M.E., 2016. Urban Agglomeration and Local Economic Growth in China: The Role of New Industrial Parks (SSRN Scholarly Paper No. ID 2746711). Social Science Research Network, Rochester, NY.

Zhu, S., He, C., Zhou, Y., 2017. How to jump further and catch up? Path-breaking in an uneven industry space. J Econ Geogr 17, 521–545. https://doi.org/10.1093/jeg/lbw047

Zhu, S., Wang, C., He, C., 2019. High-speed Rail Network and Changing Industrial Dynamics in Chinese Regions. International Regional Science Review 42, 495–518. https://doi.org/10.1177/0160017619835908

Zhu, S., Yu, C., He, C., 2020. Export structures, income inequality and urban-rural divide in China. Applied Geography 115, 102150. https://doi.org/10.1016/j.apgeog.2020.102150